\documentclass[12pt]{article}
\usepackage{amsmath,amssymb,amsfonts}
\usepackage{graphicx}

\usepackage[english]{babel}
\usepackage{bm}
\usepackage[utf8]{inputenc}
\usepackage{listings}
\usepackage[T1]{fontenc}
\usepackage{lmodern}
\usepackage{caption}
\usepackage{multirow}
\usepackage{soul}
\usepackage{orcidlink} 
\usepackage[normalem]{ulem} 
\usepackage{floatrow}
\usepackage{subcaption}
\usepackage{mathrsfs}
\usepackage{enumerate}
\usepackage[normalem]{ulem}
\usepackage{color}
\usepackage{array}
\usepackage{csquotes}

\usepackage{xcolor}  
\usepackage[table]{xcolor}

\usepackage[
backend=biber,
style=numeric-comp,
sorting=none
]{biblatex}
\begin{document}
\title{JUNO's Impact on the Neutrino Mass Ordering from Lorentz Invariance Violation}

\date{}
\author{
Tatiana Araya-Santander$\orcidlink{0009-0003-8328-3834}$\footnote{E-mail: {\tt tatiana.araya@alumnos.ucn.cl}},~\,
Cesar Bonilla$\orcidlink{0000-0002-4450-5946}$\footnote{E-mail: {\tt cesar.bonilla@ucn.cl }},~
\, \text{and}\,
Supriya Pan$\orcidlink{0000-0003-3556-8619}$\footnote{E-mail: {\tt supriya.pan@ucn.cl}}
\\[3mm]
	\textit{ Departamento de F\'isica, Universidad Cat\'olica del Norte,}\\
	\textit{Avenida Angamos 0610, Antofagasta, 1240000, Chile}\\[3mm]
    }

\maketitle

\begin{abstract}
\noindent
We explore the potential of the Jiangmen Underground Neutrino Observatory (JUNO) to probe new physics by searching for Lorentz-invariance violation (LIV). Using the 59.1-day dataset recently released by this experiment, we analyze neutrino oscillations within the minimal Standard Model Extension framework, focusing on isotropic CPT-even ($c_{ee}$$-$$c_{e\mu}$, $c_{ee}$$-$$c_{e\tau}$) and CPT-odd ($a_{ee}$$-$$a_{e\mu}$, $a_{ee}$$-$$a_{e\tau}$) parameter combinations. Our analysis of JUNO data reveals a significant shift in the oscillation parameter space of $\sin^2\theta_{12}$$-$$\Delta m^2_{21}$ when LIV is included, with the best-fit point for normal ordering moving to higher values of the solar angle $\theta_{12}$ with respect to the standard case. This creates a substantial separation between the $3\sigma$ regions of normal and inverted orderings that is absent in the standard scenario. We also find that the inverted ordering yields a lower $\chi^2_{\min}$ than the normal ordering in the LIV scenarios considered. The $c_{ee}$$-$$c_{e\tau}$ and $a_{ee}$$-$$a_{e\tau}$ sectors show the most pronounced effects. From the resulting $\chi^2$ profiles, we derive phenomenological constraints on these LIV parameter combinations using the recent JUNO data release. These bounds provide complementary information to existing limits and illustrate JUNO's sensitivity to physics beyond the Standard Model.
\end{abstract}

\maketitle

\section{Introduction}\label{sec:intro}
The Standard Model (SM) describes most known interactions in particle physics, but it still cannot explain several important observations. A prominent example is neutrino oscillation, which demonstrates that neutrinos possess non-zero masses and undergo flavor mixing —features absent in the minimal formulation of the SM. In the three-flavor framework, neutrino propagation in vacuum leads to oscillation probabilities of the form \cite{PhysRevD.110.030001} 

\begin{eqnarray}
\label{eq:nuprob}
P_{\nu_\alpha \rightarrow \nu_\beta}(L,E) &=& \delta_{\alpha\beta} 
- 4 \sum_{i>j} \text{Re}\left(U_{\alpha i}^* U_{\beta i} U_{\alpha j} U_{\beta j}^*\right) \sin^2\left(\frac{\Delta m_{ij}^2 L}{4E}\right)\notag \\
&+& 2 \sum_{i>j} \text{Im}\left(U_{\alpha i}^* U_{\beta i} U_{\alpha j} U_{\beta j}^*\right) \sin\left(\frac{\Delta m_{ij}^2 L}{2E}\right),
\end{eqnarray}

where $U_{\alpha i}$ are the elements of the Pontecorvo-Maki-Nakagawa-Sakata (PMNS) mixing matrix, $\Delta m_{ij}^2 \equiv m_i^2- m_j^2$ denotes the mass-squared differences, $L$ is the distance between the source and detector, and $E$ is the neutrino energy.

The PMNS matrix is parameterized by three mixing angles ($\theta_{12}, \theta_{13}, \theta_{23}$), a Dirac CP-violating phase $\delta_{\mathrm{CP}}$, and -if neutrinos are Majorana particles- two additional Majorana phases that do not affect oscillation probabilities. Oscillations are thus governed by these three mixing angles, two independent mass-squared differences ($\Delta m^{2}_{21}, \Delta m^{2}_{31}$), and $\delta_{\mathrm{CP}}$.

While $\theta_{12}$, $\theta_{13}$, $\Delta m^{2}_{21}$, and $|\Delta m^{2}_{31}|$ are known with good precision, several fundamental questions remain. They are, the value of $\delta_{\mathrm{CP}}$, the octant of $\theta_{23}$, and the sign of $\Delta m^{2}_{31}$, which defines the neutrino mass ordering. The normal ordering (NO) corresponds to $m_3 > m_2 > m_1$, whereas the inverted ordering (IO) implies $m_2 > m_1 > m_3$. Addressing these questions, along with determining the Dirac/Majorana nature of neutrinos and their absolute mass scale, drives current experimental efforts worldwide, utilizing neutrinos from radioactive elements (Xenon-136, Germanium-76, Calcium-48), accelerator beams, nuclear reactors, as well as atmospheric, solar, and astrophysical neutrinos.

Experimental sensitivity to oscillation parameters is governed by the ratio $L/E$, which determines the oscillation frequency. Reactor antineutrino experiments, operating at MeV-scale energies with baselines from $\mathcal{O}(10)$ to $\mathcal{O}(10^5)$ meters \cite{PhysRevD.110.030001}, are particularly sensitive to $\theta_{12}$, $\theta_{13}$, $\Delta m^{2}_{21}$, and $\Delta m^{2}_{31}$ through precise measurements of the $\bar{\nu}_e$ survival probability.

The Jiangmen Underground Neutrino Observatory (JUNO) exemplifies this approach as a next-generation, high-precision reactor neutrino experiment. With excellent energy resolution and background rejection capabilities, JUNO aims to precisely measure solar parameters and determine the neutrino mass ordering (MO). Although its initial 59.1-day dataset is insufficient for a MO determination, it already provides the world's most precise constraints: $\sin^{2}\theta_{12} = 0.3092 \pm 0.0087$ and $\Delta m^{2}_{21} = (7.50 \pm 0.12) \times 10^{-5}~\mathrm{eV}^{2}$ (assuming NO) \cite{abusleme2025first}.

This high precision also makes JUNO an ideal facility for probing subleading effects from new physics beyond the SM. In this work, we explore JUNO's sensitivity to Lorentz invariance violation (LIV). Lorentz invariance underpins the isotropy and homogeneity of local relativistic quantum field theories, including the minimal SM. Its violation would signal that the SM is a low-energy limit of a more fundamental theory. We compute JUNO's sensitivity to both CP-conserving and CP-violating LIV scenarios using its initial data release.

Several neutrino experiments have previously examined LIV effects. Constraints on LIV parameters have been obtained from accelerator \cite{Majhi:2019tfi, KumarAgarwalla:2019gdj, Rahaman:2021leu, Majhi:2022fed, Fiza:2022xfw, Pan:2023qln, Agarwalla_2023_JHEP07_216, Delgadillo:2024vqu, Cordero:2024hjr,  giarnetti2024, Bora:2025xfj}, atmospheric \cite{sahoo2022probing, Raikwal:2023lzk,  SAHOO2023137949, Hennig:2025dgh}, and solar neutrino data \cite{SNO:2018mge}. More recently, reactor neutrino oscillations have also been investigated in the context of LIV \cite{Lin:2025aym}. Additional studies have considered LIV in the interpretation of high-energy astrophysical neutrino events, particularly the event KM3-230213A observed at KM3NeT \cite{Cattaneo:2025uxk, Yang:2025kfr, Satunin:2025uui}, as well as in neutrinoless double-beta decay experiments \cite{EXO-200:2016hbz, KATRIN:2022qou}. Recent works in \cite{Capozzi:2025ovi, Goswami:2025wla} analyze existing standard neutrino oscillation data in combination with JUNO data. Our work extends this program by performing a dedicated analysis of JUNO's capabilities to constrain CPT-even/odd LIV coefficients using its recently released real data. In this work, we focus on isotropic Standard Model Extension (SME) coefficients, which do not lead to sidereal modulations but instead modify the energy dependence of the reactor antineutrino oscillation spectrum. These effects are quantified through the $\chi^2$ profiles obtained from the spectral fit.

This article is organized as follows. Section~\ref{sec:LIV} outlines the theoretical framework for Lorentz invariance violation within neutrino oscillations. Section~\ref{sec:JUNO_anal} provides a description of the JUNO experiment, its capabilities relevant to this analysis, and details our numerical methodology and statistical procedures. The resulting constraints on LIV parameters and their physical implications are discussed in Section~\ref{sec:results}. Our conclusions are presented in  Section~\ref{sec:conclusion}.

\section{Lorentz Invariance Violation in Neutrino Oscillation}
\label{sec:LIV}

Following the motivation outlined in the Introduction, we present the theoretical framework for Lorentz invariance violation in neutrino oscillations. While the standard three-flavor oscillation paradigm successfully explains a wide range of experimental data, several unresolved anomalies and theoretical considerations motivate the search for new physics beyond this minimal framework \cite{PhysRevD.69.016005}. In particular, Planck-scale physics ($m_P \simeq 10^{19} \ \rm GeV$) could induce tiny low-energy violations of fundamental symmetries, such as Lorentz invariance, which might be accessible in precision neutrino experiments.

A general approach to parameterizing Lorentz and CPT violation is provided by an extension of the SM that incorporates fixed background tensor fields. These background fields couple to SM fields, inducing subtle deviations from standard particle behavior that may become observable in high-precision experiments. Within the neutrino sector, this SM extension modifies the propagation Hamiltonian by including additional terms in the Lagrangian. The contribution of these terms to the effective neutrino Hamiltonian, at leading order, is parameterized as~\cite{PhysRevD.58.116002, PhysRevD.70.031902}

\begin{equation}
\mathcal{L}_{\rm LIV} \supset
- (a_L)_\mu \bar{\psi}\gamma^\mu\psi
+ \frac{1}{2} i (c_L)_{\mu\nu}\bar{\psi}\gamma^\mu\overset{\leftrightarrow}{D}{}^{\nu}\psi + \dots,
\end{equation}

where $(a_L)_\mu$ and $(c_L)_{\mu\nu}$ are matrices in flavor space. The coefficients $(a_L)_\mu$ are CPT-odd and therefore break both CPT and Lorentz invariance, while the coefficients $(c_L)_{\mu\nu}$ are  CPT-even and break Lorentz invariance but preserve CPT. When applied to neutrinos, the leading contributions from these operators take the form

\begin{equation}
\mathcal{L}^{\nu}_{\rm eff}
= \mathcal{L}^{\nu}_{\rm SM}
- (a_L)^\mu_{\alpha\beta}\bar{\nu}_\alpha\gamma^\mu\nu_\beta
+ \frac{1}{2} i (c_L)^{\mu\nu}_{\alpha\beta}
\bar{\nu}_\alpha \gamma_\mu \overset{\leftrightarrow}{\partial_\nu}\nu_\beta.
\end{equation}

Here, $\alpha, \beta = e, \mu, \tau$ denote flavor indices, while $\mu, \nu = 0,1,2,3$ are spacetime indices. These terms modify neutrino propagation by introducing new contributions to the effective Hamiltonian. Including standard vacuum oscillations, matter effects, the time-like LIV matrix components ($\mu, \nu = 0$) in the Sun-centered isotropic reference frame, the full Hamiltonian relevant for oscillations is as follows,

\begin{equation}
H_{\rm eff} = H_0 + H_{\rm MSW} + H_{\rm LIV},
\end{equation}

where $H_{\rm MSW}$ is the contribution due to standard matter interactions, and $H_{\rm LIV}$ stands for LIV contribution given by
\begin{align}
\label{eq:H-liv}
H_{\rm LIV} &= (a_L)_{\alpha\beta}
- \frac{4}{3} E_\nu (c_L)_{\alpha\beta}  \notag \\ &=\begin{bmatrix}
		a_{ee}&a_{e\mu}&a_{e\tau}\\
		a^\star_{e\mu}&a_{\mu\mu}&a_{\mu\tau}\\
		a^\star_{e\tau}&a^\star_{\mu\tau}&a_{\tau\tau}\end{bmatrix}-
	\frac{4}{3}E_\nu \begin{bmatrix}
		c_{ee}&c_{e\mu}&c_{e\tau}\\
		c^\star_{e\mu}&c_{\mu\mu}&c_{\mu\tau}\\
		c^\star_{e\tau}&c^\star_{\mu\tau}&c_{\tau\tau}
	\end{bmatrix}.
\end{align}

The factor $-4/3$ appears\footnote{As a rotational invariant coordinate with isotropic conditions was chosen, the factor $-4/3$ appears due to that property of CPT-even coefficients. The derivation can be found in \cite{sahoo2022probing}.} because the trace component of $(c_L)$ is not observable in oscillation experiments \cite{barenboim}. The key feature of (\ref{eq:H-liv}) is the distinct energy dependence of the two terms: CPT-odd coefficients $a_L$ produce energy-independent shifts to the Hamiltonian, whereas CPT-even coefficients $c_L$ introduce 
corrections that grow linearly with the neutrino energy. As a result, different types of experiments—reactor, solar, atmospheric, and accelerator— are sensitive to different combinations of SME parameters.

In the isotropic limit considered in this work, the SME coefficients correspond to time-independent modifications of the effective neutrino Hamiltonian in the Sun-centered reference frame. Therefore, they do not produce sidereal variations in the event rate. Their expected signature in a reactor experiment is instead an energy-dependent modification of the $\bar{\nu}_e$ survival probability, which can translate into subtle changes in the reconstructed prompt-energy spectrum. In this analysis, these effects are not identified through a visual excess in the spectrum, but through changes in the $\chi^2$ profiles obtained from the spectral fit. The resulting bounds should therefore be interpreted as spectral constraints on isotropic SME coefficients within this framework.

Since the LIV-induced changes are expected to be small, the sensitivity depends on the precision with which energy-dependent deviations from the standard oscillation pattern can be resolved. High statistics, good energy resolution, and controlled systematic uncertainties are therefore important for probing such subleading effects. Medium-baseline reactor experiments offer a clean environment for this type of spectral test. In the next section, we introduce the JUNO experiment and describe the numerical framework used to quantify its sensitivity to isotropic LIV coefficients.

\section{JUNO: Experimental Setup and Analysis}
\label{sec:JUNO_anal}
The Jiangmen Underground Neutrino Observatory is a multi-purpose medium-baseline neutrino experiment located in China. It consists of a 20 kton liquid scintillator detector situated about 52.5 km from the Yangjiang and Taishan nuclear power plants. The primary goal of JUNO is to determine the neutrino MO with a significance of 3-4$\sigma$ by measuring the energy spectrum of reactor antineutrinos with unprecedented precision \cite{10.21468/SciPostPhysProc.17.020}. The medium baseline was chosen because the survival probability of electron antineutrinos is minimal. In addition to this main objective, JUNO will also provide precise 
measurements of several neutrino mixing parameters and contribute to searches for physics beyond the 
Standard Model.

JUNO is capable of observing a wide range of neutrino sources, including solar, atmospheric, geoneutrinos, and supernova neutrinos. Reactor antineutrinos are primarily detected through the inverse beta decay processes,

\begin{equation}
    \bar{\nu}_e + p \rightarrow e^+ + n.
\end{equation}

The first signal (prompt) in the detector is a scintillation signature left by the positron $e^+$ and its annihilation, producing two gamma rays of $E_\gamma = 0.511$ MeV. Then there is a second gamma signal (delayed) coming from neutron $n$ capture of $E_\gamma = 2.223$ MeV. The prompt and delayed signals are linked by $200$ $\mu$s, which acts as a veto to reject backgrounds. The neutrino energy $E_\nu$ is linked to prompt energy $E_{\text{prompt}}$ as $E_\nu \simeq E_{\text{prompt}} + 0.78 $ MeV.

It is expected that JUNO will measure the two solar parameters, $\Delta m_{21}^2$ and $\sin^2 \theta_{12}$, as well as the atmospheric mass splitting $\Delta m_{31}^2$, after six years of data-taking.

Using (\ref{eq:nuprob}), the reactor antineutrino survival probability in vacuum is given by

\begin{align}
     P_{\bar{\nu}_e \rightarrow \bar{\nu}_e}& = 1 - \cos^2\theta_{13} \sin^22\theta_{12} \sin^2\left(\frac{\Delta m_{21}^2 L}{4E}\right) \notag \\
     & \quad - \sin^22\theta_{13} \left[\cos^2 \theta_{12}\sin^2\left(\frac{\Delta m_{31}^2 L}{4E} \right) + \sin^2 \theta_{12}\sin^2\left(\frac{\Delta m_{32}^2 L}{4E} \right) \right],
     \label{eq:Peeprob}
\end{align}

where $\Delta m^2_{ij} \equiv m_i^2-m_j^2$ is the squared neutrino mass difference, $L$ the source--detector distance, and $E$ is the neutrino energy.

The first term, proportional to $\Delta m_{21}^2$ and $\theta_{12}$, corresponds to the `solar' oscillation. It has a long oscillation length and produces a wide valley in the probability. The second term, driven by $\Delta m_{31}^2$ and $\theta_{13}$, generates oscillations with a much shorter period (`rapid oscillations') and appears as small ripples superimposed on that valley \cite{10.21468/SciPostPhysProc.17.020}.

In presence of LIV the reactor antineutrino survival probability becomes~\cite{Lin:2025aym}

\begin{align}
     P_{\bar{\nu}_e \rightarrow \bar{\nu}_e}& = 1 - \cos^2\theta_{13} \sin^22\theta_{12} \sin^2\left(\frac{\Delta m_{21}^2 L}{4E} + f_{21}\right) \notag \\
     & \quad - \sin^22\theta_{13} \left[\cos^2 \theta_{12}\sin^2\left(\frac{\Delta m_{31}^2 L}{4E}+ f_{31} \right) + \sin^2 \theta_{12}\sin^2\left(\frac{\Delta m_{32}^2 L}{4E} + f_{31} \right) \right],
     \label{eq:PeeprobLIV}
\end{align}

where the frequencies $f_{ij}$ (with $i,j=1,2,3$) are defined in~\cite{Lin:2025aym}.

Notice that the oscillation probability in (\ref{eq:Peeprob}) changes in presence of LIV, resulting in (\ref{eq:PeeprobLIV}), where the oscillation frequencies are modified by the LIV dependent parameters $f_{21}$ and $f_{31}$~\cite{Lin:2025aym}. Since the experimental data depend on these oscillation frequencies, the inclusion of LIV can affect the fit through $f_{21}$ and $f_{31}$. However, this analytical expression for the probability is not used in our numerical analysis. To assess JUNO's sensitivity to LIV effects, we perform a quantitative study using the recently released 59.1-day data set \cite{abusleme2025first}. The simulations are carried out with the GLoBES framework \cite{Huber_2005, Huber_2007, Kopp2010snu1.0}, which we modify to include the LIV Hamiltonian described in Section \ref{sec:LIV}.

The statistical analysis follows the standard $\chi^2$ formalism with pull terms. The total $\chi^2$ is obtained by marginalizing over the oscillation parameters $\omega$ and the systematic pull variables $\xi$, as follows

\begin{equation}
\label{eq:chi-tot}
    \chi^2 = {\rm min} \left[\chi^2_{\rm stat}(\omega,\xi) + \chi^2_{\rm pull}(\xi) \right].
\end{equation}

The Poisson statistical contribution is defined as 

\begin{equation}
    \chi^2_{\rm stat}(\omega,\xi)
    = 2\sum_i \left[ N_i^{\rm test} - N_i^{\rm true} 
    + N_i^{\rm true}\,\ln\!\left(\frac{N_i^{\rm true}}{N_i^{\rm test}}\right) \right],
\end{equation}

while the pull term encoding systematic uncertainties is given by

\begin{equation}
    \chi^2_{\rm pull} = \sum_{r=1}^4 \xi_r^2 \, ,
\end{equation}

where the four pull parameters correspond to the two signal systematic uncertainties for each of the two reactor contributions, Yangjiang and Taishan; one associated with the signal normalization and one with the spectral calibration. 

Here, $N_{i}^{\rm true}$ denotes the observed number of events in the $i$th JUNO energy bin of the 59.1-day dataset, while $N_{i}^{\rm test}$ represents the predicted number of events obtained from the theoretical model under test implemented in GLoBES. The ranges over which the oscillation parameters are varied are listed in Table~\ref{tab1}.

\begin{table}[h]
\centering
\begin{tabular}{|c|c|}
\hline
Parameter & Variation range \\ 
\hline
$\theta_{12}$      & $[30.0^\circ,\, 36.1^\circ]$ \\
$\theta_{13}$      & $[8.3^\circ,\, 8.6^\circ]$ \\
$\Delta m^2_{21}$  & $[7.1,\, 8.1]\times 10^{-5}\,\text{eV}^2$ \\
$\Delta m^2_{31}$  & $[2.4,\, 2.5]\times 10^{-3}\,\text{eV}^2$ \\
\hline
\end{tabular}
\caption{Variation ranges of standard oscillation parameters used in the analysis. The parameters $\theta_{23}$ and $\delta_{\rm CP}$ do not affect the reactor antineutrino oscillation probability and are therefore not varied. We are using the current bounds on $\theta_{13}$ and $\Delta m^2_{31}$ from the Daya Bay experiment \cite{PhysRevLett.130.161802}.}
\label{tab1}
\end{table}

\begin{figure}[h!]
    \centering
    \includegraphics[width=0.46\linewidth]{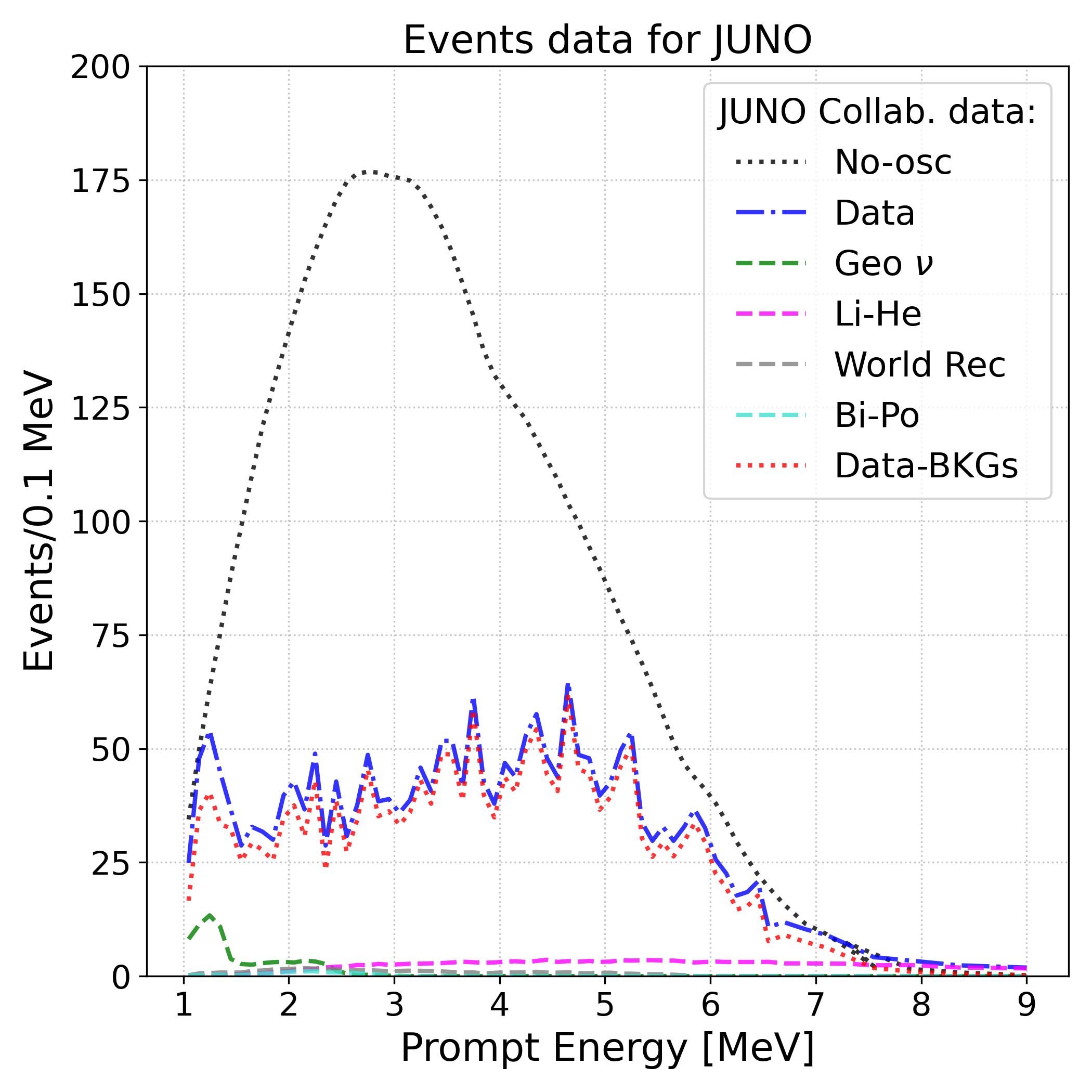}
    \includegraphics[width=0.46\linewidth]{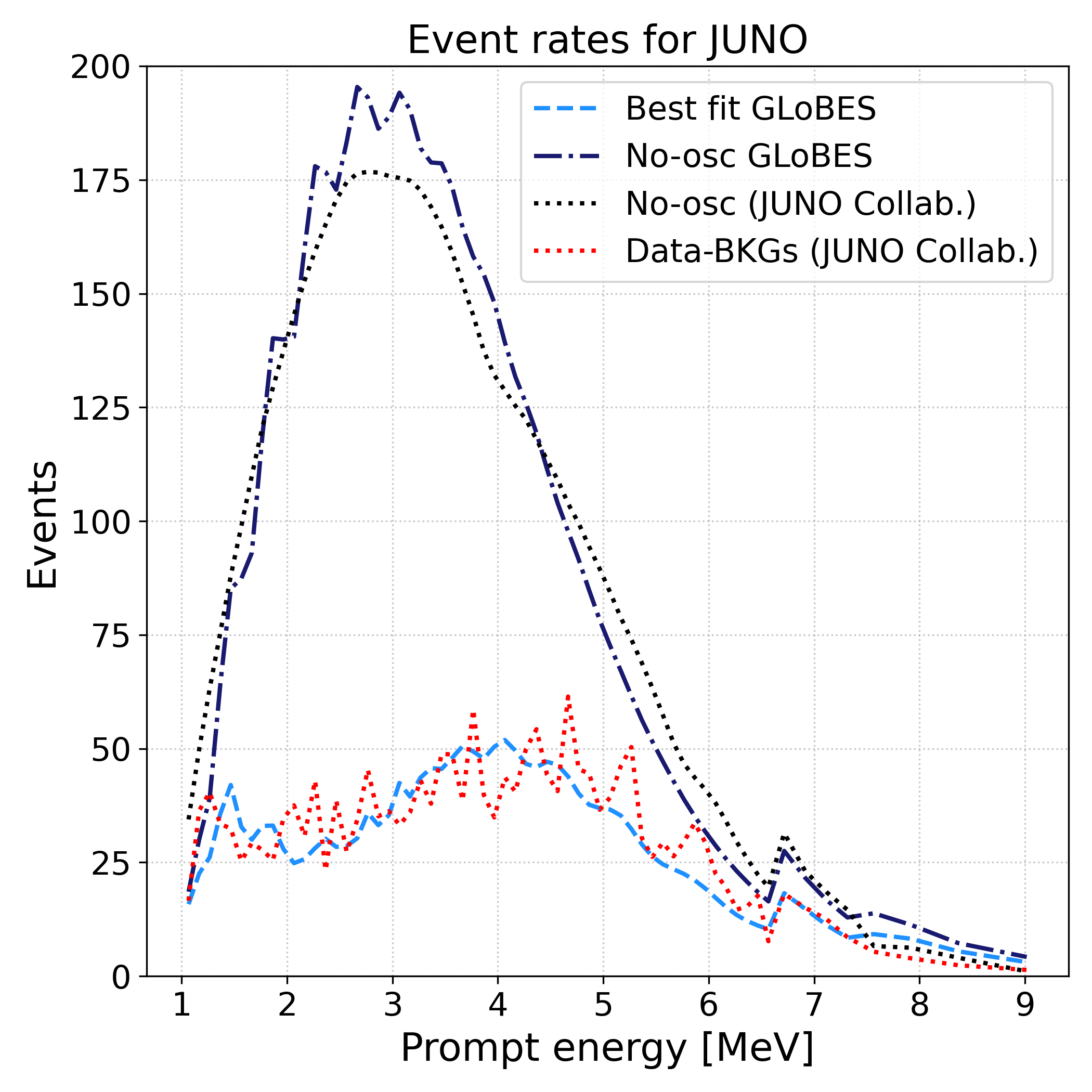}
    \caption{Left: Reconstructed JUNO prompt energy spectrum (per 0.1 MeV) from 59.1 day data set \cite{abusleme2025first}. The blue curve shows the raw experimental data, while the red curve represents the data after subtracting all background components, shown individually as the green (geo-$\nu$), pink ($^{9}$Li/$^{8}$He), grey (world reactors), and light blue ($^{214}$Bi-$^{214}$Po) contributions. The black dotted curve represents the expected unoscillated reactor flux. Right: Unoscillated event rates from JUNO (black) compared with GLoBES simulations (dark blue). The events using the best-fit obtained using GLoBES prediction for standard oscillation for NO are shown in light blue. The experimental events after subtracting all the backgrounds from JUNO data (blue) are shown by the red curve in both panels. These comparisons validate the input used for the LIV sensitivity analysis.}
    \label{fig:juno-data}
\end{figure}

In our analysis, we first normalize the unoscillated events obtained from GLoBES to match the unoscillated JUNO spectrum shown by the black dotted curve in both panels and by the dark blue curve in the right panel of Figure \ref{fig:juno-data}. The observed event numbers $N_i^{\rm true}$ are taken from the publicly released JUNO reconstructed prompt-energy spectrum after background subtraction, while the predicted event numbers $N_i^{\rm test}$ are generated using GLoBES, including possible LIV contributions.

Since the released JUNO spectrum is given at reconstructed-energy level, detector response effects are already included in the experimental data. To compare with it, the GLoBES prediction includes the finite energy resolution of the detector through the energy-resolution function implemented in the AEDL file, with a  JUNO-like resolution parameter of $3\%$. Statistical uncertainties are included through the $\chi^2$ calculation, and systematic uncertainties are treated with GLoBES pull method, including $0.8\%$ signal-normalization uncertainty and a $0.5\%$ spectral-calibration uncertainty. The signal efficiency is implemented separately through an overall signal factor of $0.8$. Since the full JUNO covariance matrix is not used, the resulting bounds should be interpreted within this statistical and detector response treatment.

The analysis uses 64 prompt-energy bins spanning the range $E_{\rm prompt}$ in $[1.0, 9.4]$ MeV, i.e. $[1.8, 10.2]$ MeV for neutrino energy $E_\nu$, assuming a baseline of $52.5$ km and an energy resolution of $3\%$. The first 56 bins in the range $[1.0, 6.6]$ MeV have a width of $0.1$ MeV, the next four bins in the range $[6.6, 7.4]$ MeV have a width of $0.2$ MeV, followed by bins of $0.3$ MeV, $0.5$ MeV, and $0.8$ MeV, respectively. 

To validate our numerical implementation, we compare the unoscillated and oscillated spectra obtained within GLoBES with those reported by the JUNO collaboration \cite{abusleme2025first}. The total number of unoscillated events reproduced in our setup agrees with the JUNO expectation. While minor differences in the spectral shape are visible near the peak region, the position of the oscillation features in energy is consistently reproduced, which is essential for the $\chi^2$ scans presented below. Moreover, the best-fit point obtained in the standard three-neutrino scenario lies close to the JUNO Collaboration result, showing that our independent GLoBES implementation provides a consistent description of the released spectrum for the purposes of the sensitivity study performed here.

There is an apparent enhancement in the event spectrum around 6.6 MeV in the right panel of the Figure \ref{fig:juno-data}. This feature originates from the fact that the right panel shows event counts per bin, whereas the left panel displays the JUNO spectrum normalized to a uniform bin width of $0.1$ MeV (events/0.1 MeV). The relation between both representations is
\begin{equation}
    \rm events/bin = events/0.1 \, MeV \times \frac{bin \, width}{0.1 \, MeV}.
\end{equation}

In particular, in the range of $[6.6, 9.4]$ MeV, the last eight bins cover larger energy intervals, leading to an accumulation of events and an apparent enhancement in that region. Since the $\chi^2$ calculation in GLoBES is performed in terms of events per bin, the right panel displays the relevant quantity used in the statistical analysis.

With this validated numerical setup, we proceed to quantify JUNO’s sensitivity to isotropic LIV coefficients.

\section{Results}
\label{sec:results}

We first test our GLoBES implementation in the standard three neutrino oscillation scenario. Using the background subtracted JUNO spectrum as input, we compute the $\Delta\chi^2$ profiles in the $\sin^2\theta_{12}$$-$$\Delta m^2_{21}$ plane as in \cite{abusleme2025first}.

\begin{figure}[H]
    \centering
    \includegraphics[width=0.46\linewidth]{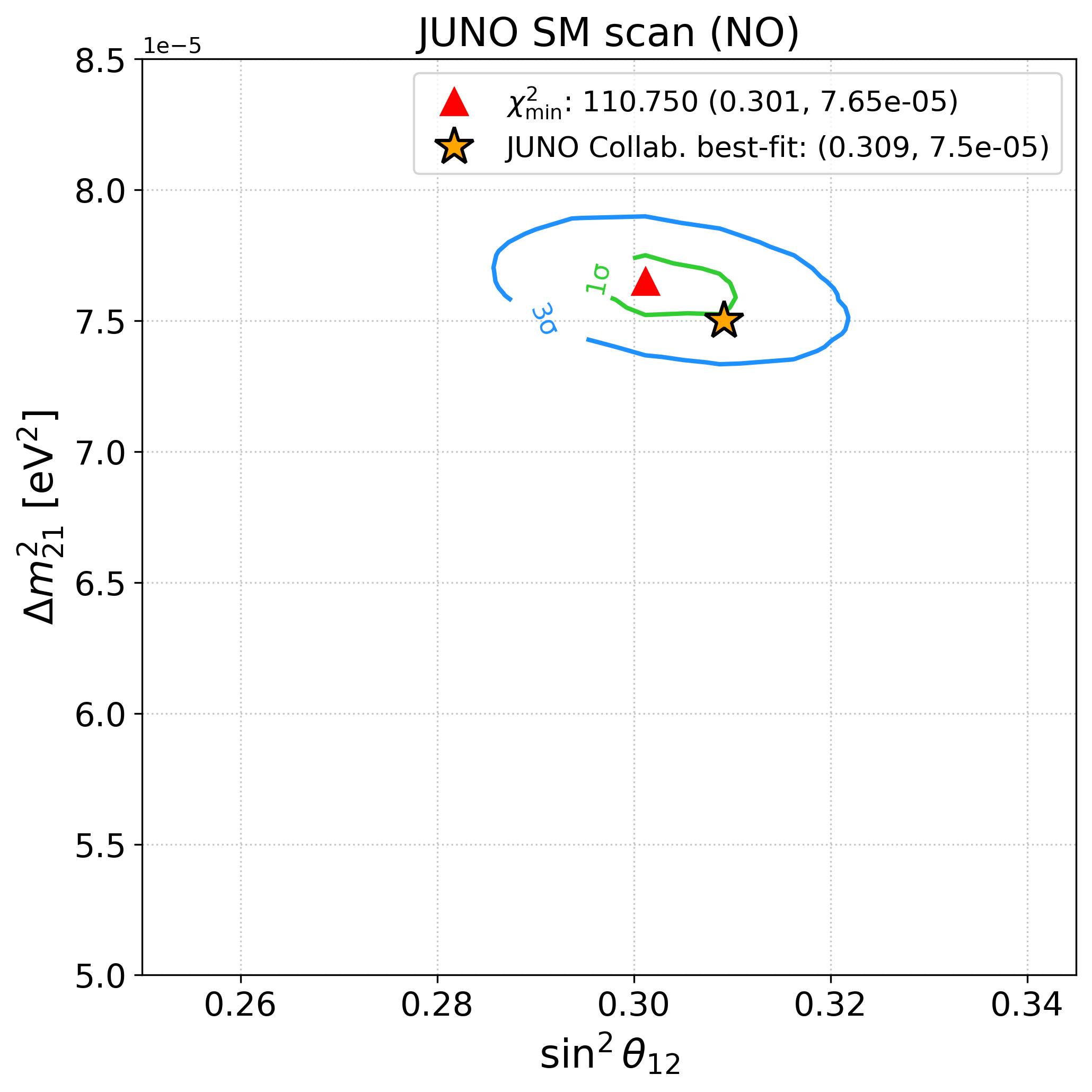}
    \includegraphics[width=0.46\linewidth]{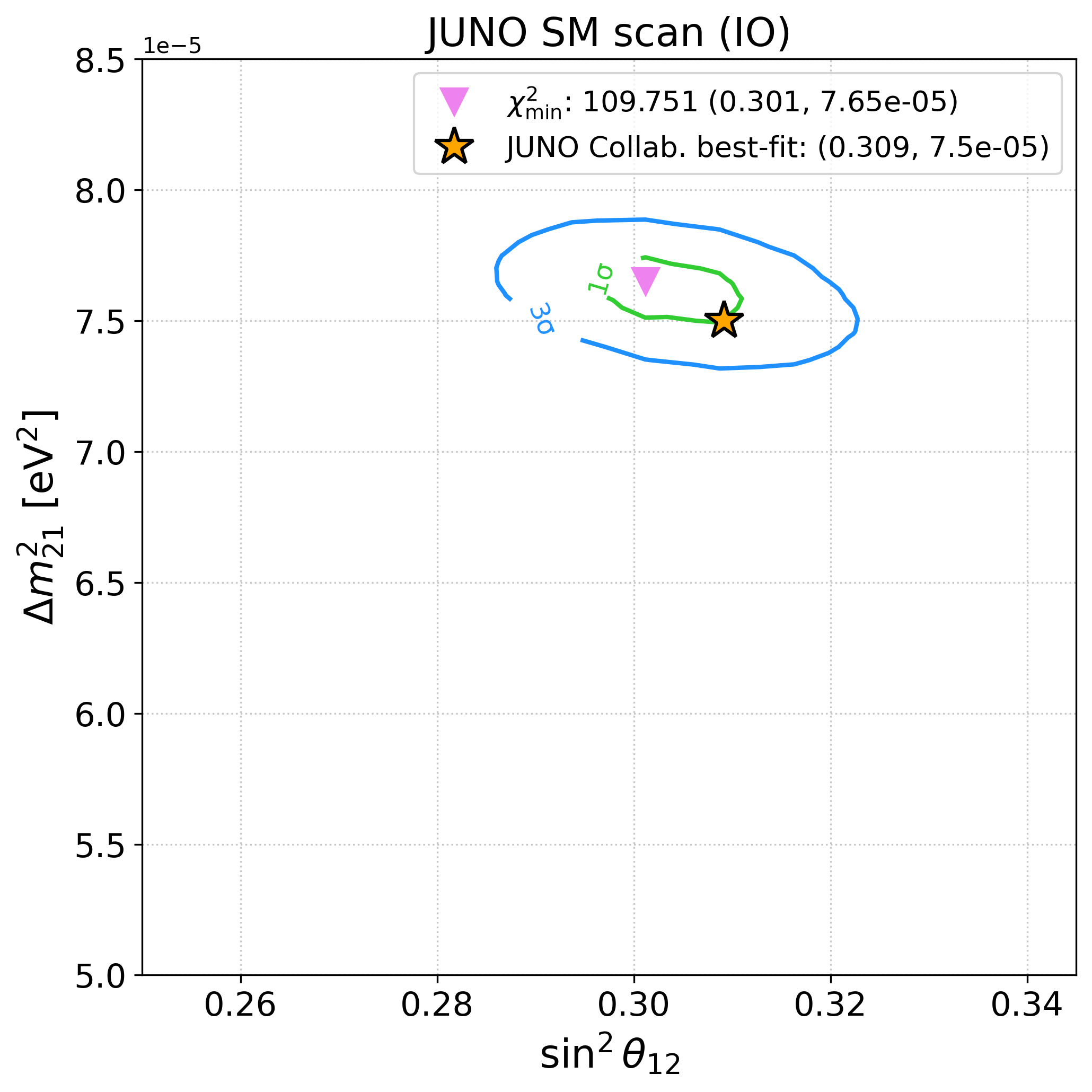}    
    \caption{Total chi-square in the $\sin^2\theta_{12}$$-$$\Delta m_{21}^2$ plane for NO (left) and IO (right), obtained from our JUNO SM scan after marginalizing over $\Delta m^2_{31}$ and $\theta_{13}$. The green and blue contours correspond to the $1\sigma$ and $3\sigma$ allowed regions, respectively. The triangle denotes the minimum found in our scan. The orange star marks the JUNO Collaboration best-fit point, $(\sin^2\theta_{12},\Delta m^2_{21})=(0.3092, 7.50\times10^{-5}\,\mathrm{eV}^2)$ \cite{abusleme2025first}.}
    \label{fig:chi2_d21-th12_SM_NO}
\end{figure}

We determined the sensitivity to the solar parameters ($\theta_{12}$ and $\Delta m^2_{21}$) by marginalizing over $\theta_{13}$ and $\Delta m^2_{31}$, as detailed in Table \ref{tab1}, considering both normal ordering (NO) and inverted ordering (IO). The resulting $\chi^2$ surfaces and confidence regions in the $\sin^2\theta_{12}$$-$$\Delta m_{21}^2$ plane are shown in Figure \ref{fig:chi2_d21-th12_SM_NO}. The minimum $\chi^2$ values reveal that IO provides a slightly better fit to the JUNO data with $\chi^2_{\rm min} = 109.751$ for IO (violet triangle), compared to $\chi^2_{\rm min} = 110.750$ for the NO (red triangle). This corresponds to $\Delta\chi^2_{\rm NO-IO} =0.999$, indicating only a mild preference for IO in the standard case. 
Furthermore, the best-fit point reported by the JUNO Collaboration, $(\sin^2\theta_{12},\Delta m^2_{21})=(0.309,\,7.5\times10^{-5}\,\mathrm{eV}^2)$~\cite{abusleme2025first}, is shown in Figure \ref{fig:chi2_d21-th12_SM_NO} for comparison. This point lies within the $3\sigma$ allowed region obtained in our scan and is close to our best-fit point. Overall, this provides a consistency check of the standard oscillation setup used before introducing LIV effects.

We now examine the sensitivity of JUNO data to LIV effects. We first focus on the CPT-even coefficients, which correspond to the CP-conserving, energy-dependent sector defined in (\ref{eq:H-liv}).

The sensitivity is explored through two-dimensional projections of the LIV parameter space, including the coefficient-pair scans $(c_{ee},c_{e\mu})$ and $(c_{ee},c_{e\tau})$, as well as the coefficient-phase scans $(c_{e\mu},\phi_{e\mu})$ and $(c_{e\tau},\phi_{e\tau})$. The analysis assumes normal ordering, while the corresponding inverted ordering results are shown in Appendix~\ref{secA1}. In all following scenarios, the standard oscillation parameters are marginalized within the ranges listed in Table \ref{tab1}. 

\begin{figure}[H]
    \centering
    \includegraphics[width=0.46\linewidth]{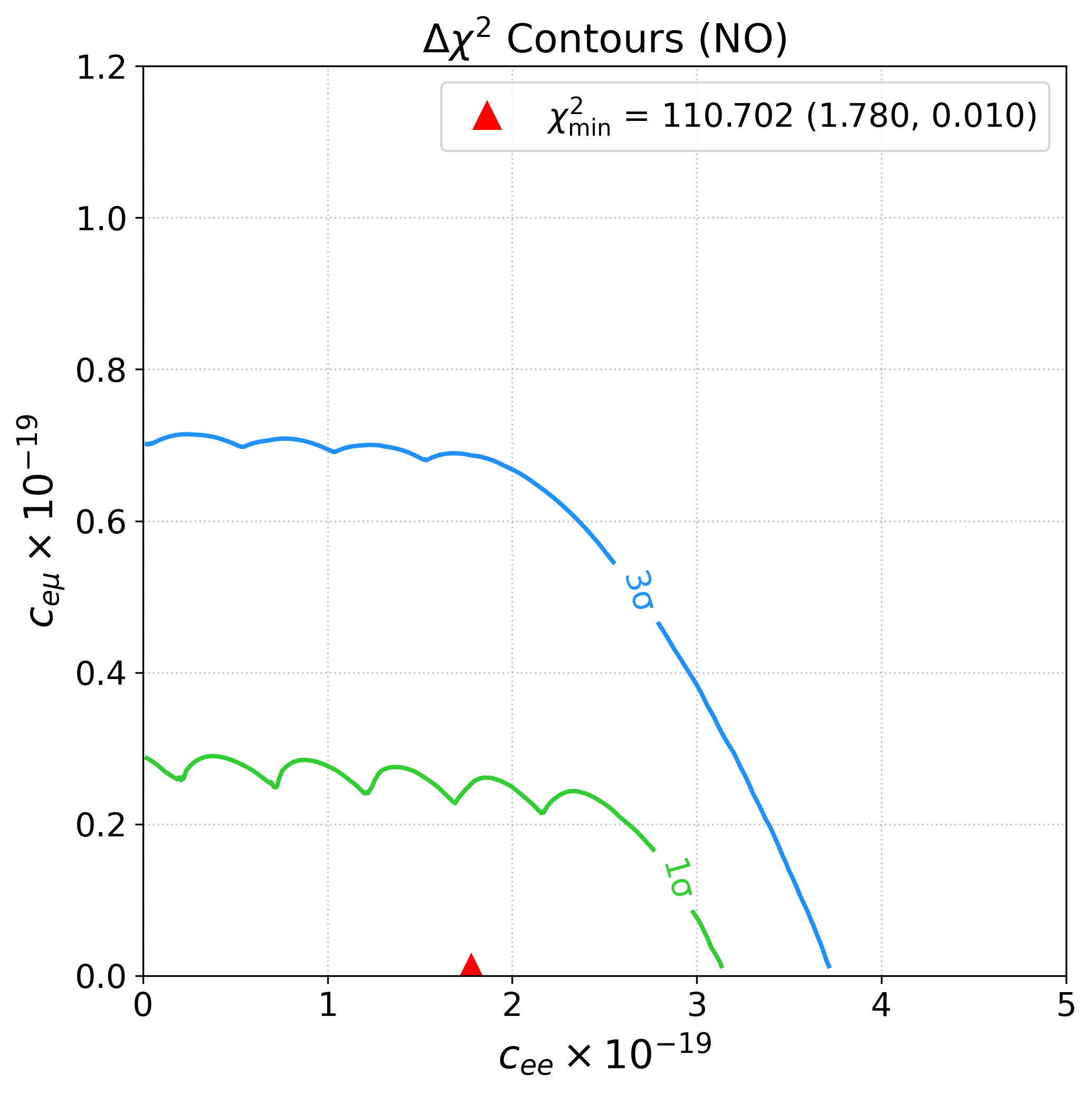}
    \includegraphics[width=0.46\linewidth]{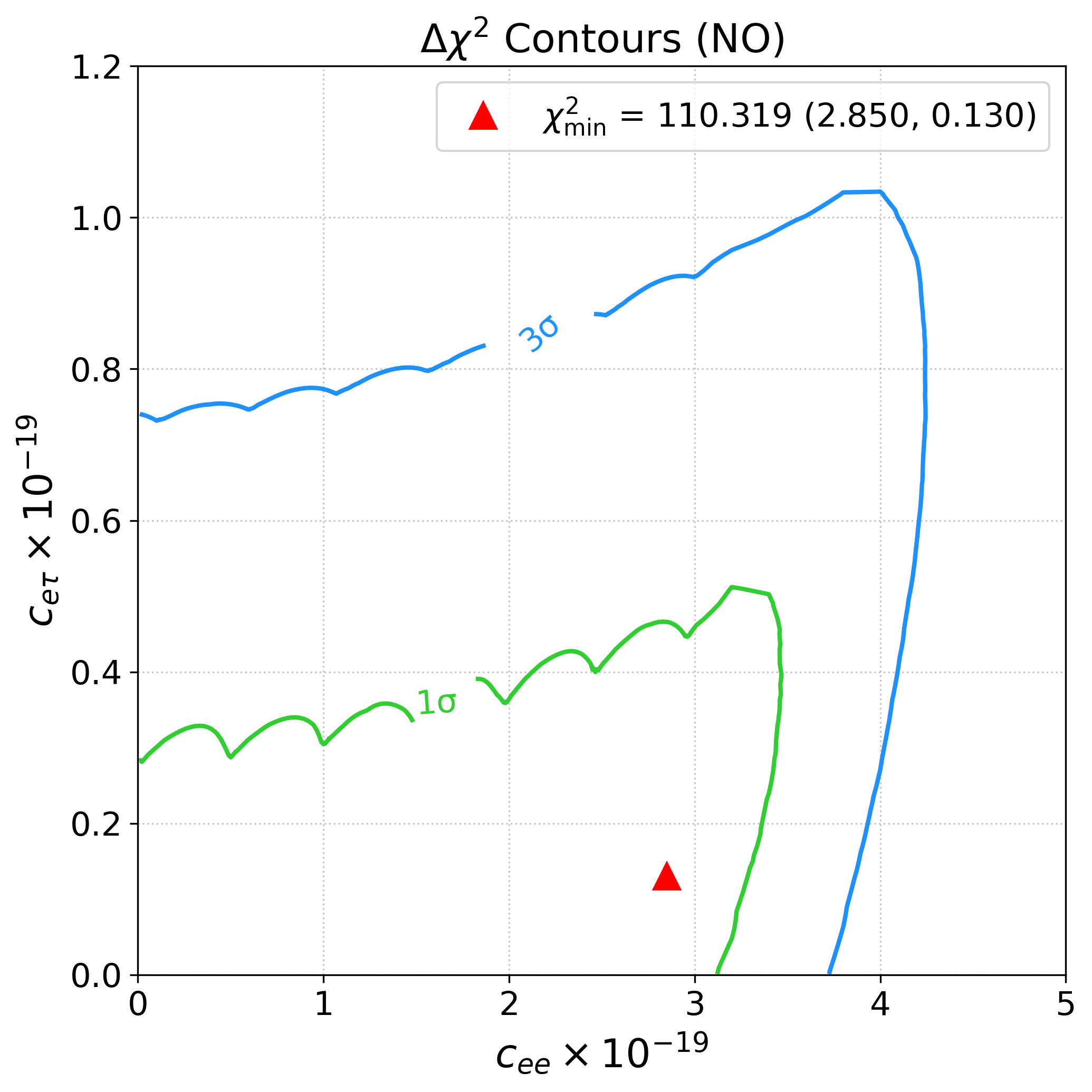}
    \caption{Two dimensional $\Delta\chi^{2}$ contours for the CP-conserving parameter pairs $(c_{ee},c_{e\mu})$ (left) and $(c_{ee},c_{e\tau})$ (right) assuming normal ordering. The red triangles mark the global minima of each scan, which correspond to the best-fit LIV values preferred by the data. The green and blue contours correspond to the $1\sigma$ and $3\sigma$ allowed regions, respectively.}
    \label{fig:chi-sqr-cee-emet-no}
\end{figure}

Figure~\ref{fig:chi-sqr-cee-emet-no} shows the first set of CPT-even scans, corresponding to the $c_{ee}$$-$$c_{e\mu}$ and $c_{ee}$$-$$c_{e\tau}$ planes in the left and right panels, respectively. The coefficients are varied in the range $[10^{-22},10^{-18}]$. The main observations can be summarized as follows,

\begin{itemize}
    \item In the $c_{ee}$$-$$c_{e\mu}$ plane (left panel), values larger than $c_{ee}>3.7\times10^{-19}$ and $c_{e\mu}>0.7\times10^{-19}$ are excluded at the $3\sigma$ confidence level. The best-fit point is found at $c_{ee}\simeq 1.78\times10^{-19}$ and $c_{e\mu}\simeq 0.01\times10^{-19}$, yielding $\chi^2=110.702$, which slightly improves compared to the SM minimum $\chi^2_{\rm min}=110.750$.
    \item In the $c_{ee}$$-$$c_{e\tau}$ plane (right panel), values larger than $c_{ee}> 4.25\times10^{-19}$ and $c_{e\tau}> 1.03\times10^{-19}$ lie outside the $3\sigma$ allowed region.
    The best-fit point is obtained at $c_{ee}\simeq 2.85\times10^{-19}$, $c_{e\tau}\simeq 0.13\times10^{-19}$ with $\chi^2=110.319$ that is less than standard $\chi^2_{\rm min}=110.750$.
\end{itemize}

\begin{figure}[H]
    \centering
    \includegraphics[width=0.46\linewidth]{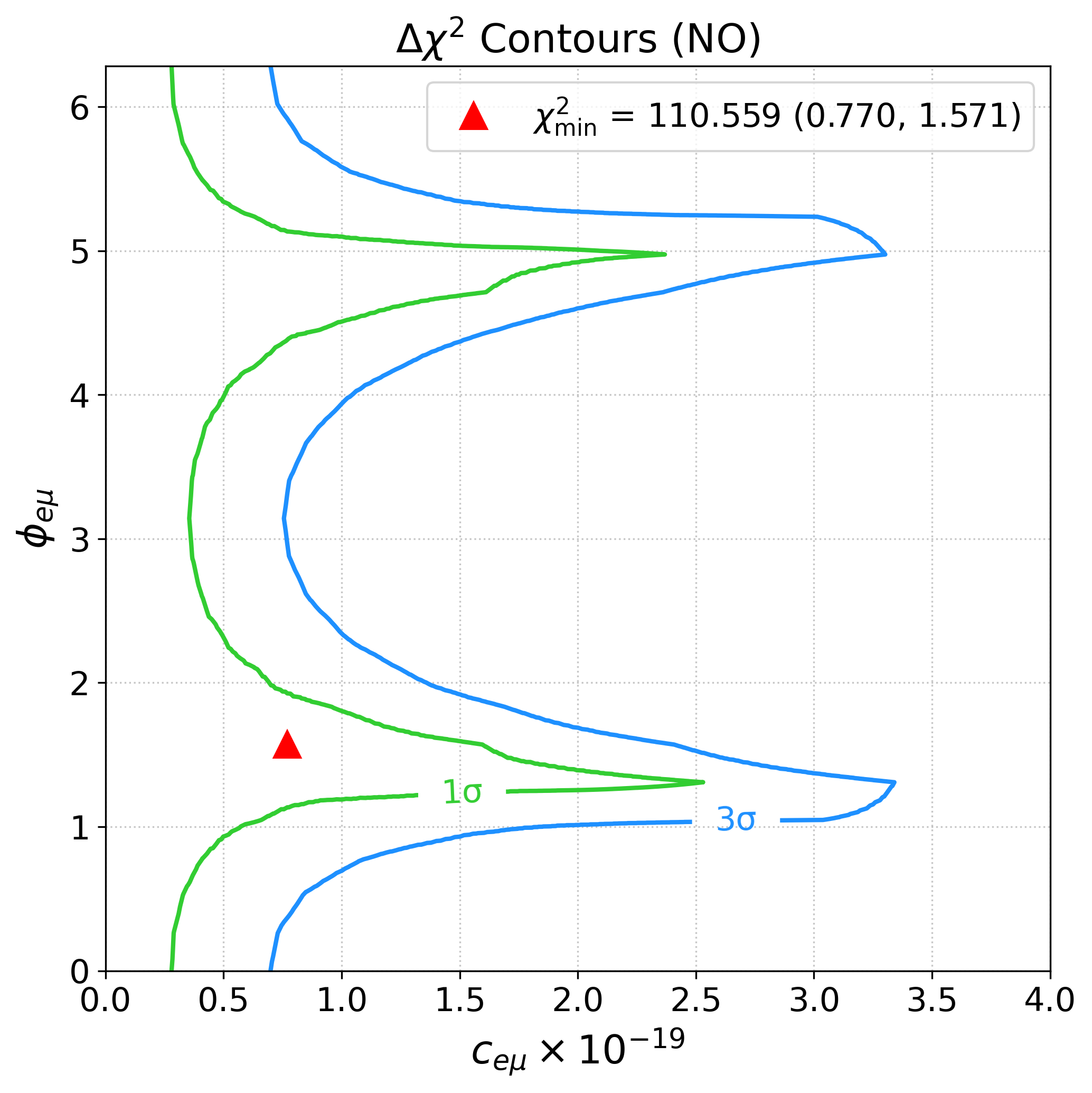}
    \includegraphics[width=0.46\linewidth]{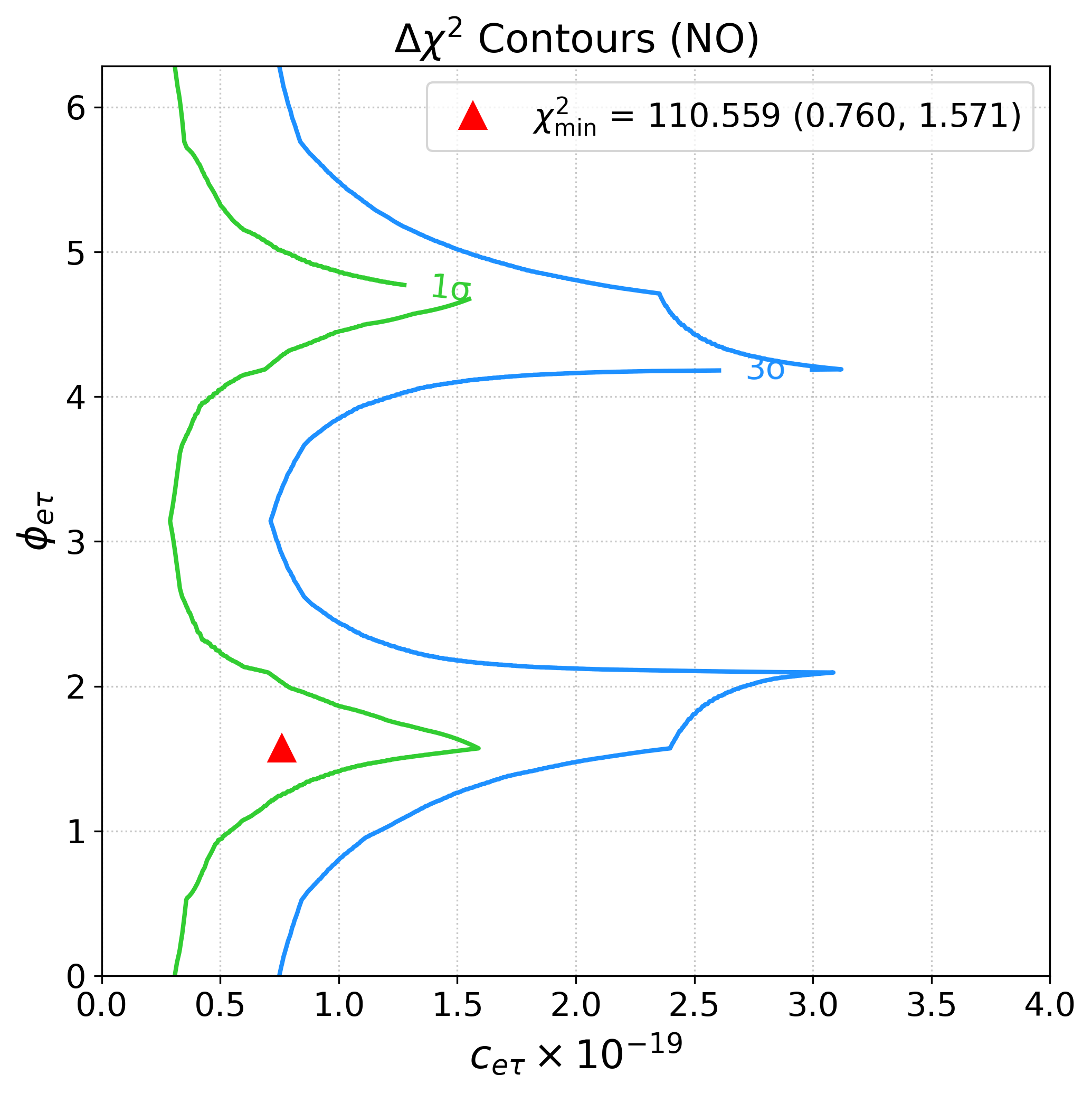}
    \caption{Two dimensional $\Delta\chi^{2}$ contours for the CP-conserving parameter pairs $(c_{e\mu}, \phi_{e\mu})$ (left) and $(c_{e\tau}, \phi_{e\tau})$ (right) assuming normal ordering. The red triangles mark the global minima of each scan, which correspond to the best-fit LIV values preferred by the data. The green and blue contours correspond to the $1\sigma$ and $3\sigma$ allowed regions, respectively.}
    \label{fig:chi-sqr-c-ph-no}
\end{figure}

Figure~\ref{fig:chi-sqr-c-ph-no} shows the second set of CPT-even scans, where one LIV coefficient is varied together with its associated phase. The left panel corresponds to $c_{e\mu}$$-$$\phi_{e\mu}$ plane, while the right panel corresponds to $c_{e\tau}$$-$$\phi_{e\tau}$ plane. The main observations are:

\begin{itemize}
    \item The phases $\phi_{e\mu}$ and $\phi_{e\tau}$ are weakly constrained, since the contours remain open over the scanned phase range.
    \item Around the phase values $\phi \simeq \pi/2$ and $3\pi/2$, the contours extend to larger values of the corresponding coefficient, meaning that the sensitivity is weaker in these regions. The approximate $3\sigma$ bounds are $c_{e\mu}\simeq 3.3\times10^{-19}$ and $c_{e\tau}\simeq 3.1\times10^{-19}$.
    \item The best-fit point in the left panel is $c_{e\mu}\simeq0.77\times10^{-19},\phi_{e\mu}=\pi/2$, while in the right panel it is $c_{e\tau}\simeq0.76\times10^{-19},\phi_{e\tau}=\pi/2$ with $\chi^2_{\rm min}=110.56$ in both cases.
\end{itemize}

The best-fit values, minimum $\chi^2$ values, and approximate $3\sigma$ bounds obtained from the CPT-even scans are summarized in Tables~\ref{tab:cpc-no} and~\ref{tab:cpc-io} for NO and IO, respectively.

\begin{table}[h]
\resizebox{\textwidth}{!}{
\begin{tabular}{|c|c|c|c|c|}
\hline
Plane & $c_{ee}$$-$$c_{e\mu}$  & $c_{ee}$$-$$c_{e\tau}$ &  $c_{e\mu}$$-$$\phi_{e\mu}$ & $c_{e\tau}$$-$$\phi_{e\tau}$\\
\hline
Best-fit & $(1.78,0.01)\times 10^{-19}$   &  $(2.85,0.13)\times 10^{-19}$  &  $(0.77\times 10^{-19},1.57)$  & $(0.76\times 10^{-19},1.57)$\\
$\chi^2_{\rm min}$ & 110.702   & 110.319  & 110.56 & 110.56 \\
$3\sigma$  & $(3.7,0.7)\times 10^{-19}$  &  $(4.25,1.03)\times 10^{-19}$  & $(3.3\times 10^{-19},$ N.A.\footnotemark[1]) & $(3.1\times 10^{-19},$ N.A.\footnotemark[1])\\
\hline
\end{tabular}
}
\vspace{0.4em}
\footnotesize
$^{1}$ The whole range is allowed.
\caption{The best-fit values, $\chi^2_{\rm min}$ at the best-fit and $3\sigma$ bounds in our analysis (Figure \ref{fig:chi-sqr-cee-emet-no}, \ref{fig:chi-sqr-c-ph-no}) for NO. A marginalization was performed on the parameter $\theta_{12}, \theta_{13}, \Delta m^2_{21}$ and $\Delta m^2_{31}$.}
\label{tab:cpc-no}%
\end{table}

\begin{table}[h]
\resizebox{\textwidth}{!}{
\begin{tabular}{|c|c|c|c|c|}
\hline
Plane & $c_{ee}$$-$$c_{e\mu}$  & $c_{ee}$$-$$c_{e\tau}$ &  $c_{e\mu}$$-$$\phi_{e\mu}$ & $c_{e\tau}$$-$$\phi_{e\tau}$\\
\hline
Best-fit & $(0.01,0.01)\times 10^{-19}$   &  $(0.01,0.05)\times 10^{-19}$  &  $(0.73\times 10^{-19},1.57)$  & $(0.72\times 10^{-19},1.57)$\\
$\chi^2_{\rm min}$ & 109.80   & 109.55  & 109.48 & 109.48 \\
$3\sigma$  & $(3.7, 0.7)\times 10^{-19}$  &  $(4.13,0.95)\times 10^{-19}$  & $(3.3\times 10^{-19},$ N.A.\footnotemark[1]) & $(3.05\times 10^{-19},$ N.A.\footnotemark[1])\\
\hline
\end{tabular}
}
\vspace{0.4em}
\footnotesize
$^{1}$ The whole range is allowed.
\caption{The best-fit values, $\chi^2_{\rm min}$ at the best-fit and $3\sigma$ bounds in our analysis (Figure \ref{fig:chi-sqr-cee-emet-io}, \ref{fig:chi-sqr-c-ph-io})  for IO. A marginalization was performed on the parameter $\theta_{12}, \theta_{13}, \Delta m^2_{21}$ and $\Delta m^2_{31}$.}
\label{tab:cpc-io}%

\end{table}

\begin{figure}[H]
    \centering
    \includegraphics[width=0.46\linewidth]{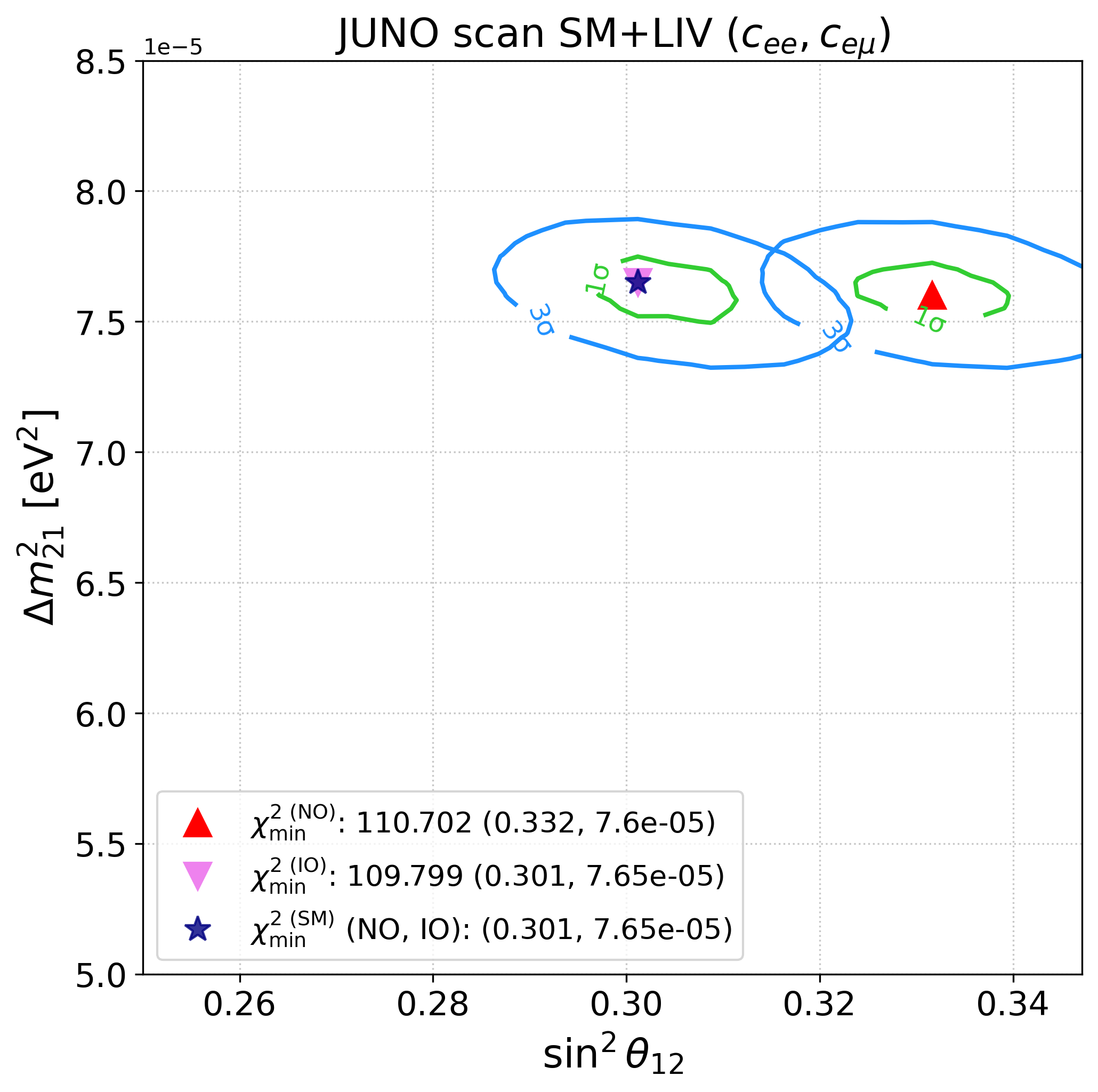}
    \includegraphics[width=0.46\linewidth]{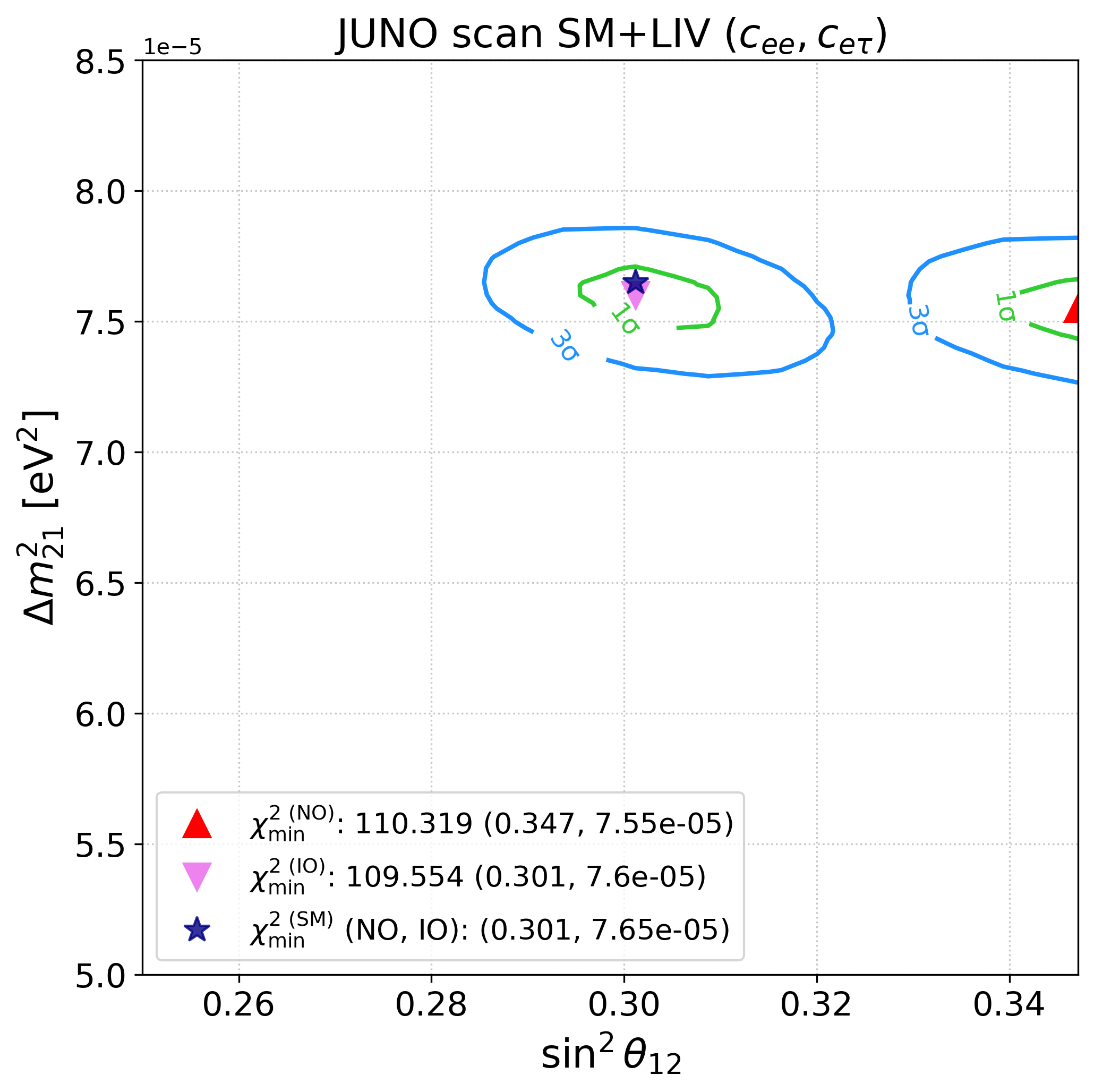}
    \caption{Sensitivity in the $\sin^2 \theta_{12}$$-$$ \Delta m_{21}^2$ plane using best-fit value of CPT-even LIV parameters $c_{ee}, c_{e\mu}$ (left) and  $c_{ee}, c_{e\tau}$ (right). The $3\sigma$ and $1\sigma$ contours of NO and IO are shown by blue and green, respectively. Best fits are pointed by red and violet triangles for NO and IO, respectively. The blue star indicates the SM best-fit point from Figure~\ref{fig:chi2_d21-th12_SM_NO}, included as a reference to show how the fit is displaced when the CPT-even LIV parameters are fixed to their best-fit values.}
    \label{fig:dm21-sin12-liv}
\end{figure}

Using these best-fit CPT-even LIV coefficients as fixed inputs, we now study how the corresponding LIV scenarios affect the solar-parameter fit. We generate JUNO+LIV sensitivity contours in the $\sin^2\theta_{12}$$-$$\Delta m^2_{21}$ plane, marginalizing over $\theta_{13}$ and $\Delta m^2_{31}$ within the ranges given in Table~\ref{tab1}. Figure~\ref{fig:dm21-sin12-liv} shows these contours for both NO and IO, in the presence of the best-fit LIV parameters $(c_{ee}, c_{e\mu})$ (left panel) and $(c_{ee}, c_{e\tau})$ (right panel). The SM best-fit point from Figure~\ref{fig:chi2_d21-th12_SM_NO} is also shown for comparison.

The main observations from Figure \ref{fig:dm21-sin12-liv} are:

\begin{itemize}
    \item In both panels, the best-fit for NO (red triangle) shifts to a higher value of $\sin^2\theta_{12}$. We find $\sin^2\theta_{12}=0.332$ for the $(c_{ee},c_{e\mu})$ case and $\sin^2\theta_{12}=0.347$ for the $(c_{ee},c_{e\tau})$. This shift moves it away from the IO best-fit (violet triangle), which remains close to $\sin^2\theta_{12}=0.301$, and the SM best-fit point represented by the blue star. As a result, the overlap between the NO and IO $3\sigma$ allowed regions is reduced.
    \item The $3\sigma$ region of IO is similar to the standard case, but it yields a lower $\chi^2_{\rm min}$ in both cases. This indicates that, within our analysis, the JUNO data prefer IO when the best-fit CPT-even LIV parameters are included. The SM best-fit point, indicated by the blue star, is also close to the IO best-fit point, supporting this behavior.
    \item For the $(c_{ee},c_{e\mu})$ case, the 3$\sigma$ contours of NO and IO partially overlap. In contrast, for the $(c_{ee},c_{e\tau})$ case, the 3$\sigma$ contours of NO are significantly distinct from IO.
\end{itemize}

\begin{figure}[H]
    \centering
    \includegraphics[width=0.46\linewidth]{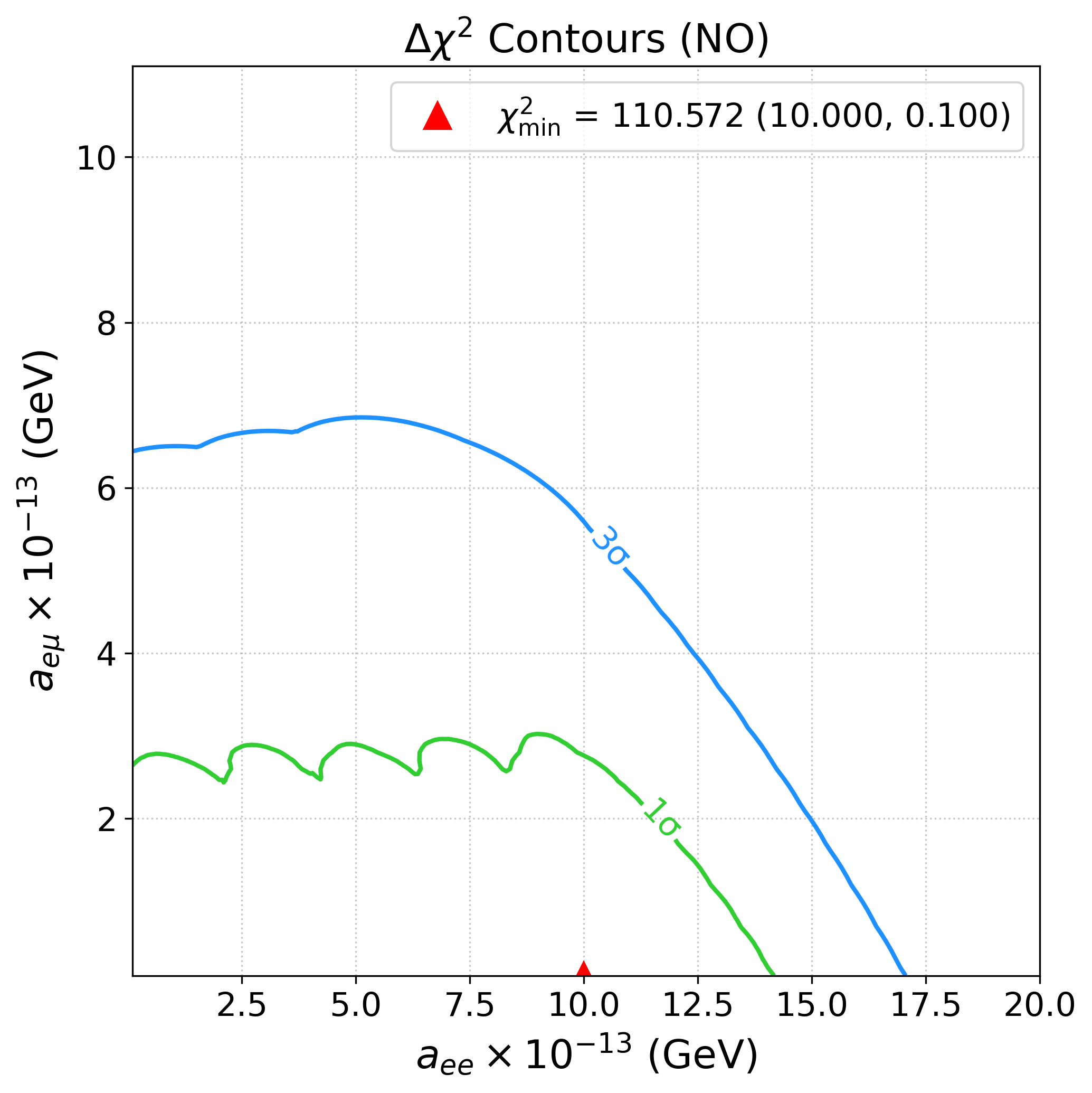}
    \includegraphics[width=0.46\linewidth]{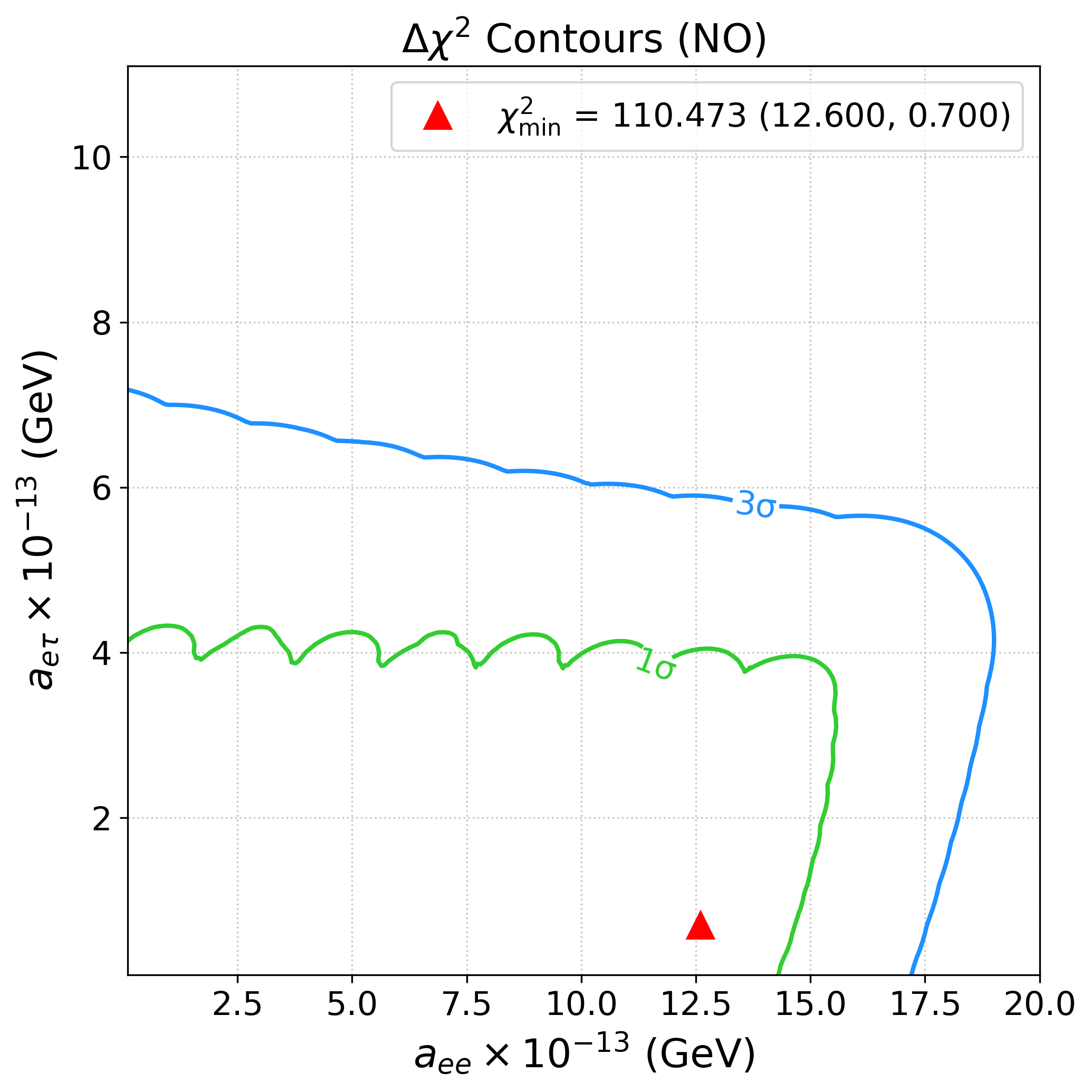}
    \caption{Two dimensional $\Delta\chi^{2}$ contours for the CP-violating parameter pairs parameter pairs $(a_{ee}, a_{e\mu})$ (left) and $(a_{ee}, a_{e\tau})$ (right) assuming normal ordering. The red triangles mark the global minima of each scan, which correspond to the best-fit LIV values preferred by the data. The green and blue contours correspond to the $1\sigma$ and $3\sigma$ allowed regions, respectively.}
    \label{fig:chi-sqr-aee-amet-no}
\end{figure}

We now shift our focus to the CPT-odd LIV sector and repeat the same strategy. The CPT-odd parameters, defined in (\ref{eq:H-liv}) by $a_{\alpha\beta}$, are independent of neutrino energy and have the dimensions of energy. Figure \ref{fig:chi-sqr-aee-amet-no}, shows the sensitivity in the $a_{ee}$$-$$a_{e\mu}$ and $a_{ee}$$-$$a_{e\tau}$ planes, shown in the left and right panels, respectively. The parameters are varied in the range $[10^{-15},10^{-12}]~\mathrm{GeV}$. The main observations are:

\begin{itemize}
    \item The $3\sigma$ bounds are stronger for $a_{e\mu}\simeq 6.8\times10^{-13}\,\mathrm{GeV}$ and $a_{e\tau}\simeq 7.1\times10^{-13}\,\mathrm{GeV}$ than for $a_{ee}\simeq 17.1$--$19\times10^{-13}\,\mathrm{GeV}$.
    \item In the $(a_{ee}, a_{e\mu})$ plane, the best-fit point is $(10.0, 0.1)\times10^{-13}$ GeV, with $\chi^2_{\rm min}=110.572$. In the $(a_{ee},a_{e\tau})$ plane, the best-fit point is $(a_{ee},a_{e\tau})=(12.60,0.7)\times10^{-13}~\mathrm{GeV}$, with $\chi^2_{\rm min}=110.473$. In both cases, the minimum is lower than the standard three-neutrino value, $\chi^2_{\rm min}=110.750$.
\end{itemize}

\begin{figure}[H]
    \centering
    \includegraphics[width=0.46\linewidth]{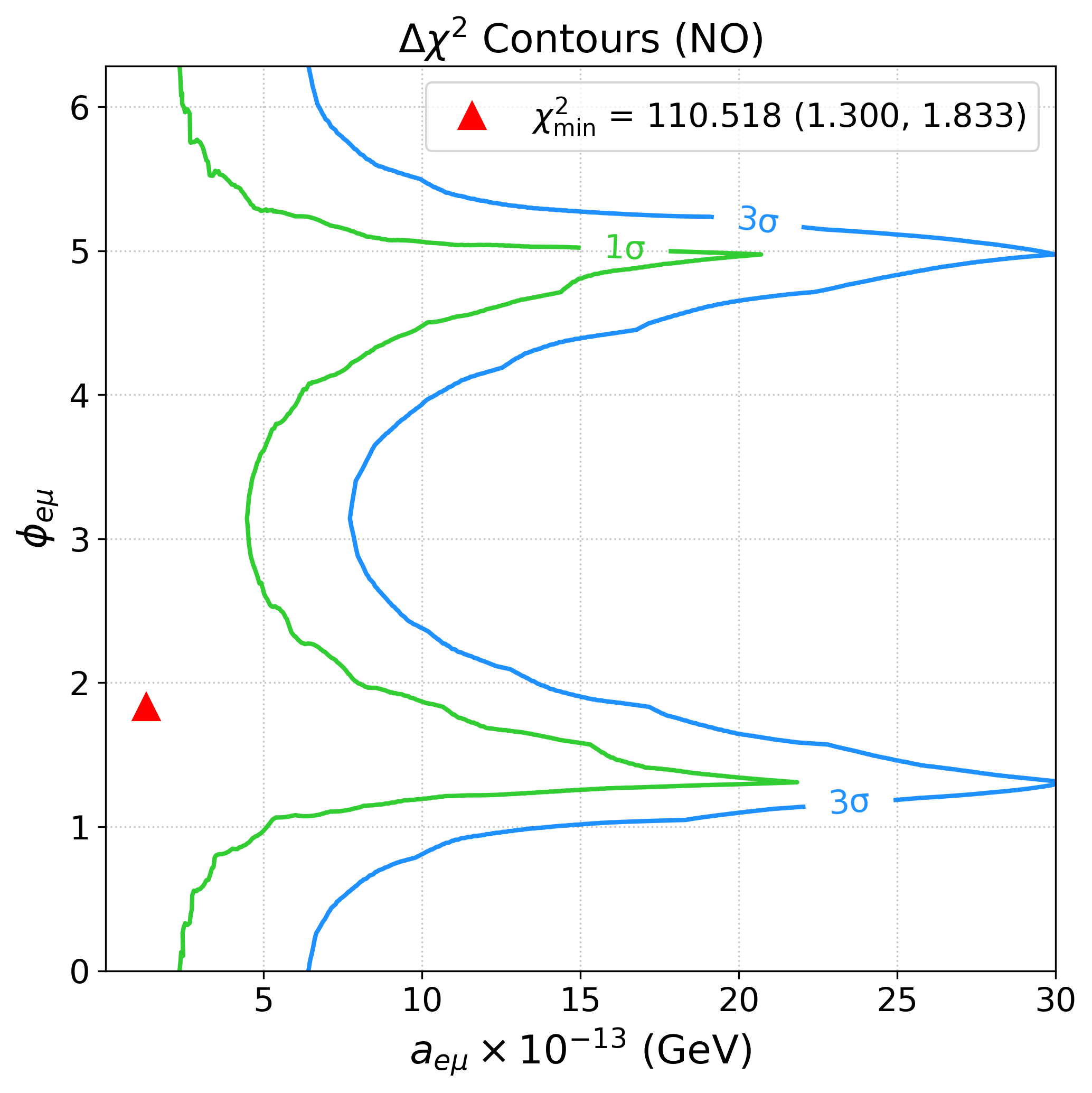}
    \includegraphics[width=0.46\linewidth]{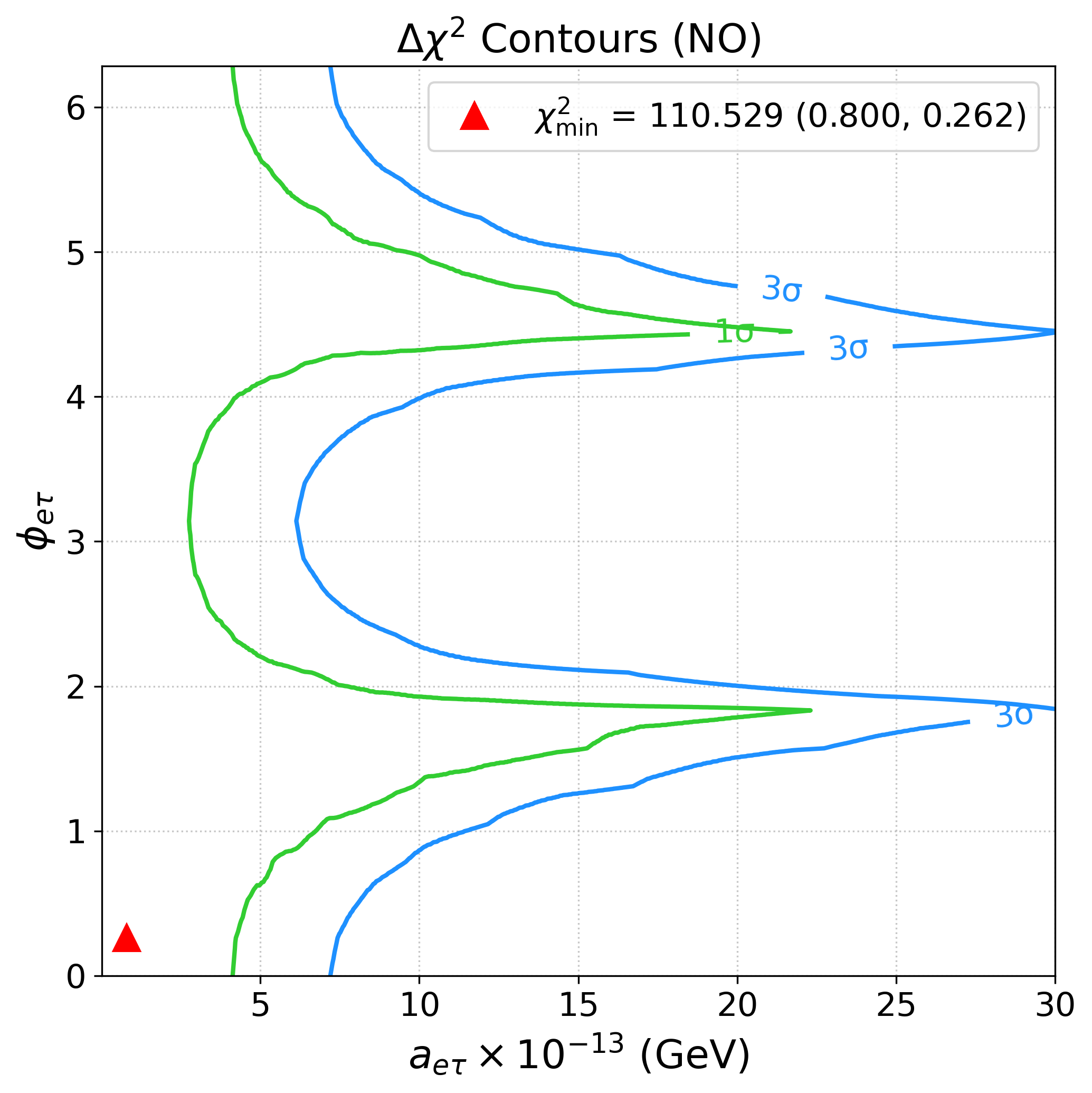}
    \caption{Two dimensional $\Delta\chi^{2}$ contours for the CP-violating parameter pairs $(a_{e\mu}, \phi_{e\mu})$ (left) and $(a_{e\tau}, \phi_{e\tau})$ (right) assuming normal ordering. The red triangles indicate the global minima of each scan, corresponding to the best-fit LIV values preferred by the data. The green and blue contours correspond to the $1\sigma$ and $3\sigma$ allowed regions, respectively.}
    \label{fig:chi-sqr-a-ph-no}
\end{figure}

As in the CPT-even sector, we also study the dependence on the phases associated with the CPT-odd coefficients. Figure~\ref{fig:chi-sqr-a-ph-no} shows the scans in the $(a_{e\mu},\phi_{e\mu})$ and $(a_{e\tau},\phi_{e\tau})$ planes, shown in the left and right panels, respectively. The behavior is similar to that found in the CPT-even phase scans of Figure \ref{fig:chi-sqr-c-ph-no}. The observations are as follows:

\begin{itemize}
    \item In both panels, the full scanned ranges of $\phi_{e\mu},\phi_{e\tau}$ are allowed, similar to CP-conserving case in Figure \ref{fig:chi-sqr-c-ph-no}.
    \item Similar to Figure \ref{fig:chi-sqr-c-ph-no}, the sensitivity to the corresponding LIV coefficients decreases around $\phi\simeq \pi/2, 3\pi/2 $, as shown by the horn-like structures in the contours.
    \item The best-fit points in both panels are different from those found in the CPT-even phase scans. For the left panel, the best-fit point is $(a_{e\mu},\phi_{e\mu})=(1.30\times10^{-13}\,\mathrm{GeV},1.83)$, whereas for the right panel it is $(a_{e\tau},\phi_{e\tau})=(0.80\times10^{-13}\,\mathrm{GeV},0.26)$. The corresponding $\chi^2$ values are lower than the standard case value.
\end{itemize}

The best-fit values, minimum $\chi^2$ values, and approximate $3\sigma$ bounds obtained from the CPT-odd scans are summarized in Tables~\ref{tab:cpv-no} and~\ref{tab:cpv-io} for NO and IO, respectively.

\begin{table}[h!]
\centering
\resizebox{\textwidth}{!}{

\begin{tabular}{ |c|c|c|c|c| } 
\hline
Plane & $a_{ee}$$-$$a_{e\mu}$  & $a_{ee}$$-$$a_{e\tau}$ &  $a_{e\mu}$$-$$\phi_{e\mu}$ & $a_{e\tau}$$-$$\phi_{e\tau}$\\
\hline
Best-fit & $(10.0, 0.1)\times10^{-13}$ GeV  & $(12.6, 0.7)\times10^{-13}$ GeV  & $(1.30\times10^{-13}$ GeV, 1.83) & $(0.80\times10^{-13}$ GeV, 0.26) \\
$\chi^2_{\rm min}$ & 110.57 & 110.47 & 110.52  & 110.53 \\
$3\sigma$  & $(17.1, 6.8)\times 10^{-13}$ GeV & $(19.0,7.1)\times 10^{-13}$ GeV & $(3.0\times10^{-12}$ GeV, N.A.\footnotemark[1]) & ($3.0\times10^{-12}$ GeV, N.A.\footnotemark[1]) \\ 
\hline

\end{tabular}
}
\footnotesize
$^{1}$ The whole range is allowed.
\caption{The best-fit values, $\chi^2_{\min}$ at the best-fit and $3\sigma$ bounds in our analysis (Figure \ref{fig:chi-sqr-aee-amet-no}, \ref{fig:chi-sqr-a-ph-no}) for NO. A marginalization was performed on the parameter $\theta_{12}, \theta_{13}, \Delta m^2_{21}$ and $\Delta m^2_{31}$.}
\label{tab:cpv-no}
\end{table}

\begin{table}[h!]
\centering
\resizebox{\textwidth}{!}{
\begin{tabular}{ |c|c|c|c|c| } 
\hline
Plane & $a_{ee}$$-$$a_{e\mu}$  & $a_{ee}$$-$$a_{e\tau}$ &  $a_{e\mu}$$-$$\phi_{e\mu}$ & $a_{e\tau}$$-$$\phi_{e\tau}$\\
\hline
Best-fit & $(1.5, 0.1 )\times10^{-13}$ GeV   & $(0.1, 0.7 )\times10^{-13}$ GeV  & $(1.30\times10^{-13}$ GeV, 4.19) & $(0.70\times10^{-13}$ GeV, 0.0) \\
$\chi^2_{\rm min}$ & 109.61   & 109.38  & 109.40  & 109.37 \\
$3\sigma$  & $(17.0,6.8)\times 10^{-13}$ GeV & $(18.9,7.2)\times 10^{-13}$ GeV & $(3.0\times10^{-12}$ GeV, N.A.\footnotemark[1]) & $(3.0\times10^{-12}$ GeV, N.A.\footnotemark[1]) \\ 
\hline
\end{tabular}
}
\footnotesize
$^{1}$ The whole range is allowed.
\caption{The best-fit values, $\chi^2_{\min}$ at the best fit and $3\sigma$ bounds in our analysis (Figure \ref{fig:chi-sqr-aee-amet-io}, \ref{fig:chi-sqr-a-ph-io}) for IO. A marginalization was performed on the parameter $\theta_{12}, \theta_{13}, \Delta m^2_{21}$ and $\Delta m^2_{31}$.}
\label{tab:cpv-io}
\end{table}

\begin{figure}[H]
    \centering
    \includegraphics[width=0.46\linewidth]{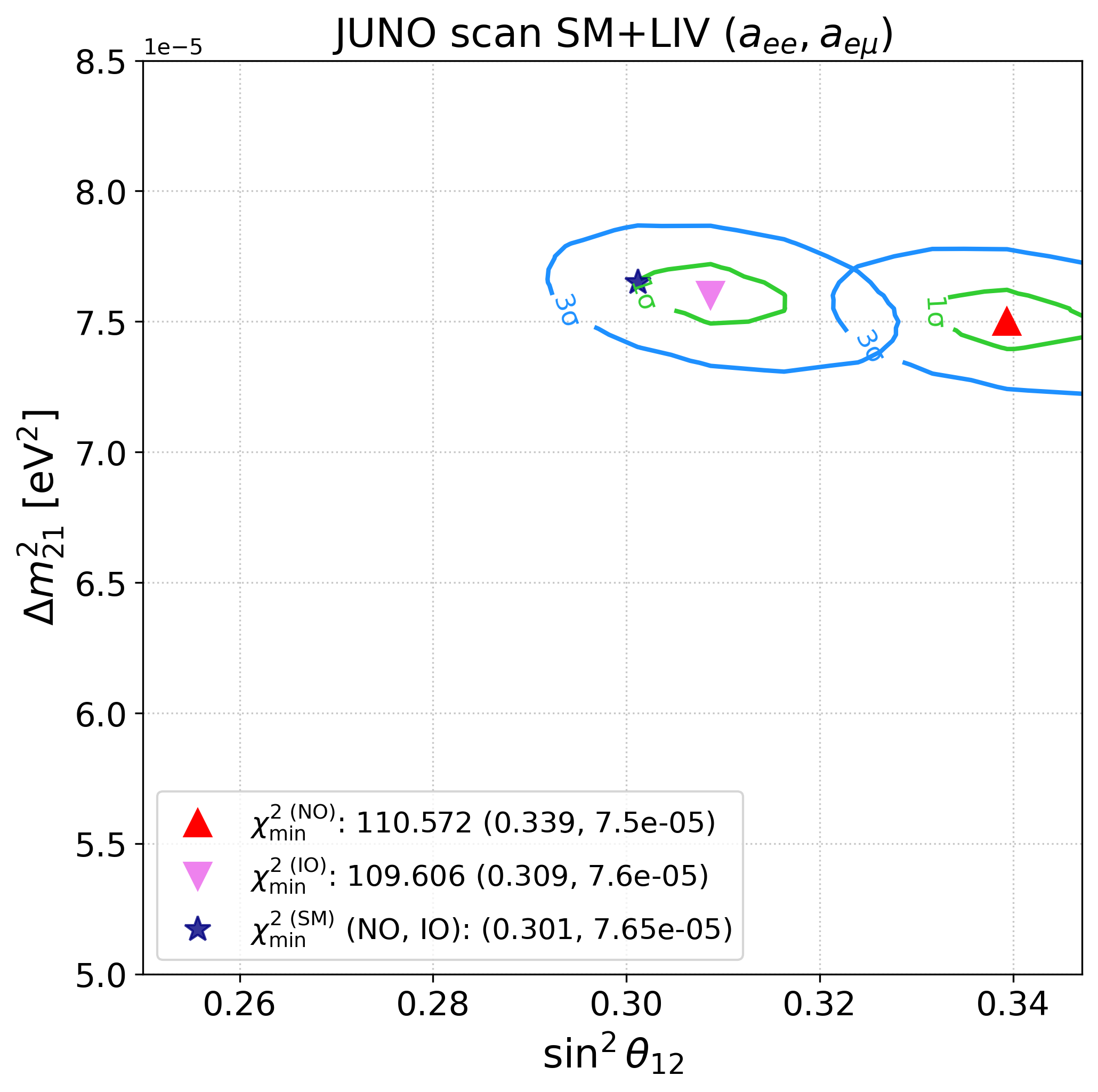}
    \includegraphics[width=0.46\linewidth]{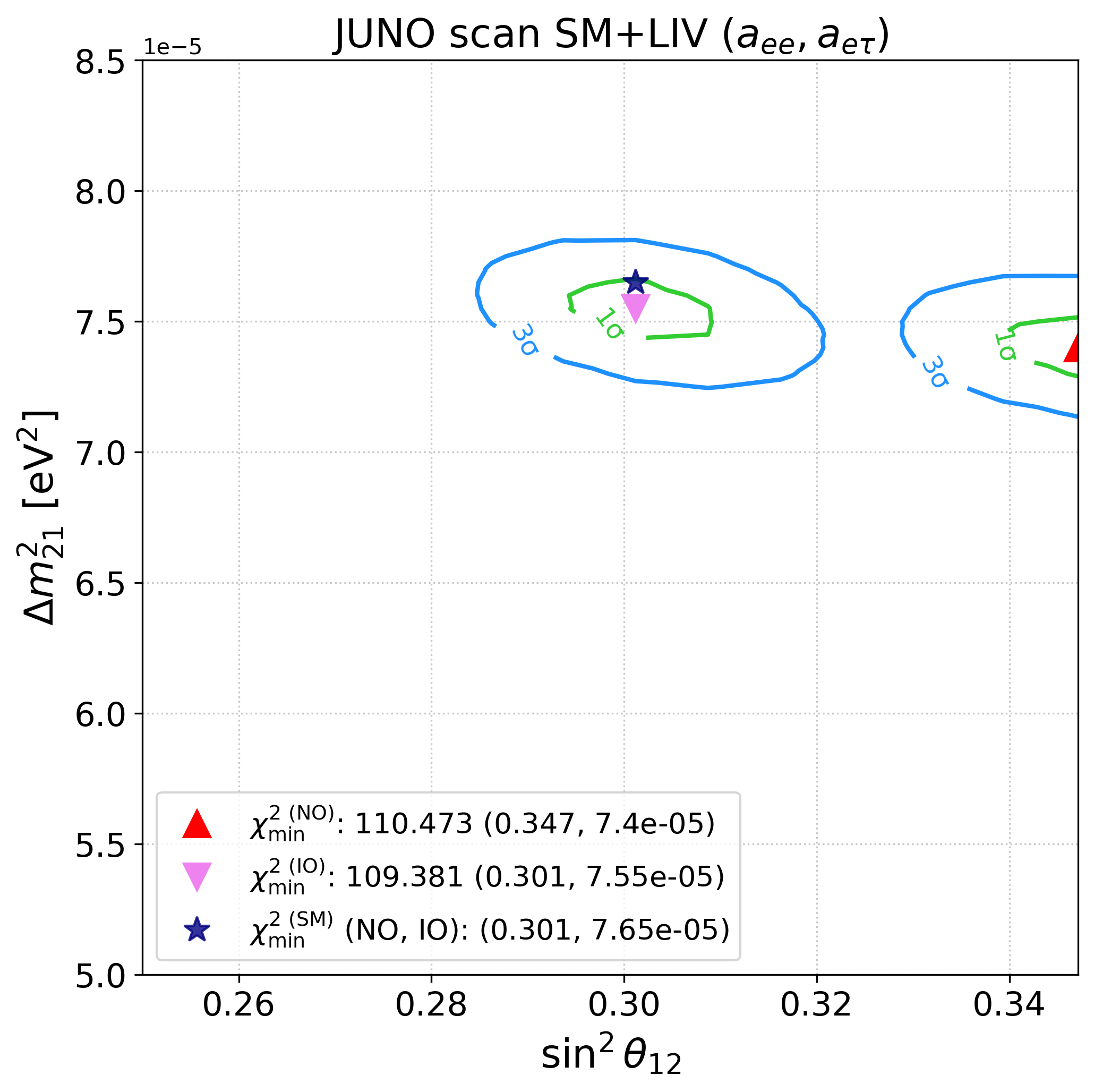}
    \caption{Sensitivity in the $\sin^2 \theta_{12}$$-$$\Delta m_{21}^2$ plane considering best-fit value of CPT-odd parameters for $a_{ee}, a_{e\mu}$ (left) and  $a_{ee}, a_{e\tau}$ (right). The $3\sigma$ and $1\sigma$ contours of NO and IO are shown by blue and green, respectively. Best fit are pointed by red and violet triangles for NO and IO, respectively. The blue star indicates the SM best-fit point of Figure~\ref{fig:chi2_d21-th12_SM_NO}, included for comparison to show how the fit is displaced when the CPT-odd LIV parameters are fixed to their best-fit values.}
    \label{fig:dm21-sin12-liv_CPTV}
\end{figure}

Following the same procedure used for the CPT-even case, we fix the CPT-odd LIV coefficients to their best-fit values and generate JUNO+LIV sensitivity contours in the $\sin^2\theta_{12}$--$\Delta m^2_{21}$ plane, marginalizing over $\theta_{13}$ and $\Delta m^2_{31}$ within the ranges given in Table~\ref{tab1}. Figure~\ref{fig:dm21-sin12-liv_CPTV} shows the resulting contours for both NO and IO, for the $(a_{ee},a_{e\mu})$ and $(a_{ee},a_{e\tau})$ best-fit configurations. The SM best-fit point from Figure~\ref{fig:chi2_d21-th12_SM_NO} is also shown for comparison.

Similar to the CPT-even cases shown in Figure~\ref{fig:dm21-sin12-liv}, the CPT-odd LIV scenarios also show a preference for IO within our $\chi^2$ framework. The principal features of Figure~\ref{fig:dm21-sin12-liv_CPTV} are,

\begin{itemize}
    \item The best-fit for NO (red triangle) shifts to larger values of $\sin^2\theta_{12}$, reaching $\sin^2\theta_{12}=0.339$ in the left panel and $0.347$ in the right panel. This shift moves it away from the IO best-fit point (violet triangle) and reduces the overlap between the NO and IO $3\sigma$ contours.
    \item The $3\sigma$ regions of IO are similar to those obtained in the standard case and also cover the SM best-fit point, which is represented by the blue star. This is closer to the IO best-fit point (violet triangle) than to the NO best-fit point (red triangle), supporting the preference for IO in the presence of the best-fit CPT-odd LIV parameters. 
    \item For the $(a_{ee},a_{e\mu})$ case, the NO and IO 3$\sigma$ contours  partially overlap. In contrast, for the $(a_{ee},a_{e\tau})$ case, the NO and IO $3\sigma$ regions are more clearly separated.
\end{itemize}

In summary, after comparing the CPT-even and CPT-odd LIV scenarios, we find that the main effect of the best-fit LIV parameters is a displacement of the NO best-fit region toward larger values of $\sin^2\theta_{12}$. In the standard oscillation fit, the two mass orderings give close minima, with $\chi^2_{\min}=110.750$ for NO and $\chi^2_{\min}=109.751$ for IO, so the present JUNO data do not distinguish between them within our SM fit. Once LIV is included, the IO region remains closer to the standard oscillation result and gives lower $\chi^2_{\min}$ values in the LIV scenarios considered, suggesting a preference for IO within our $\chi^2$ framework. This behavior is seen in Figures~\ref{fig:dm21-sin12-liv} and~\ref{fig:dm21-sin12-liv_CPTV}.

To connect these changes in the $\chi^2$ contours with the corresponding event  spectra, we show SM--LIV event rate comparisons in Appendix~\ref{sec:SM-LIV-comparison}. These spectra remain close to the SM prediction, indicating that the LIV effects are small and energy dependent. Therefore, their impact is quantified through the full $\chi^2$ minimization rather than by visual inspection of the spectrum.

Finally, we place the phenomenological bounds obtained in this work in the context of existing limits on isotropic SME coefficients in the neutrino sector. Current limits are compiled in the updated version of the Data Tables for Lorentz and CPT Violation by Kostelecký and Russell~\cite{Kosteleck__2011} (see in particular Tables~S4, D36, and D37), where bounds are typically reported under the simplifying assumption that only a single SME coefficient is nonzero at a time.

In contrast, the present analysis is based on reactor $\bar{\nu}_e$ disappearance and is therefore sensitive to isotropic coefficients involving the electron flavor, namely $a_{ee}, a_{e\mu}, a_{e\tau}$ and $c_{ee}, c_{e\mu}, c_{e\tau}$ through their impact on $P(\bar{\nu}_e\to\bar{\nu}_e)$. Our constraints are obtained from two-dimensional $\Delta \chi^2$ scans in the joint parameter space of coefficient pairs (Figures~\ref{fig:chi-sqr-cee-emet-no}--\ref{fig:chi-sqr-a-ph-no}),
and are summarized in Tables~\ref{tab:cpc-no}--\ref{tab:cpv-io}. Consequently, the reported $3\sigma$ regions should be interpreted as constraints on combinations of coefficients
rather than individual parameters.

\section{Conclusion}
\label{sec:conclusion}

We have investigated the sensitivity of the JUNO experiment to isotropic Lorentz-invariance violating effects in reactor antineutrino oscillations within the minimal Standard Model Extension framework. In this work, we focused on isotropic SME coefficients, which correspond to time-independent modifications of the effective neutrino Hamiltonian and therefore modify the energy dependence of the reactor antineutrino spectrum rather than producing sidereal variations.

A GLoBES-based simulation and statistical analysis were performed using the 59.1-day data release \cite{abusleme2025first}. We employed a GLoBES framework, incorporating both CPT-even and CPT-odd LIV contributions. We generated the oscillated event spectra across 64 prompt energy bins in the range $[1.0, 9.4]$ MeV for a 52.5 km baseline, incorporating a 3\% energy resolution. Our analysis used a Poissonian $\chi^2$ statistic with pull terms, marginalizing over the standard oscillation parameters ($\theta_{12}$, $\theta_{13}$, $\Delta m^2_{21}$, $\Delta m^2_{31}$) within their current experimental ranges. We first validated our setup by reproducing the JUNO collaboration's $\sin^2 \theta_{12}$$-$$\Delta m^2_{21}$ sensitivity contours for standard oscillations.

The analysis of this initial dataset yields numerical constraints on isotropic LIV parameter combinations, which are collected in Tables~\ref{tab:cpc-no}--\ref{tab:cpv-io}. The reported values correspond to the maximal coefficients allowed within the correlated two-dimensional $\Delta\chi^2$ scans performed in this work, with separate results obtained under the normal ordering (NO) and inverted ordering (IO) hypotheses.

In the CPT-even (CP-conserving) sector, the 3$\sigma$
scans in the $(c_{ee},c_{e\mu})$ plane yield $c_{ee} \lesssim 3.7\times10^{-19}$ and $c_{e\mu} \lesssim 0.7 \times10^{-19}$ for both mass orderings. In the $(c_{ee},c_{e\tau})$ plane, we obtain $c_{ee} \lesssim 4.25 \times10^{-19}$ for NO and $4.13 \times10^{-19}$ for IO, together with  $c_{e\tau} \lesssim 1.03 \times10^{-19}$ for NO and $0.95 \times10^{-19}$ for IO. The CPT-odd (CP-violating) sector exhibits a similar pattern. From the $(a_{ee}, a_{e\mu})$ plane, we obtain $a_{ee} \lesssim 17.1\times 10^{-13}\,\mathrm{GeV}$ for NO and $17.0\times 10^{-13}\,\mathrm{GeV}$ for IO, together with $a_{e\mu} \lesssim 6.8\times 10^{-13}\,\mathrm{GeV}$ for both mass orderings. The $(a_{ee}, a_{e\tau})$ plane yields $a_{ee} \lesssim 19.0\times 10^{-13}\,\mathrm{GeV}$ for NO and $18.9\times 10^{-13}\,\mathrm{GeV}$ for IO, together with $a_{e\tau} \lesssim 7.1\times 10^{-13}\,\mathrm{GeV}$ for NO and $7.2\times 10^{-13}\,\mathrm{GeV}$ for IO.

At the level of the marginalized 3$\sigma$ upper limits on the isotropic LIV coefficients, the results obtained under NO and IO are comparable in all scans, with $(c_{ee},c_{e\tau})$ and $(a_{ee},a_{e\tau})$ planes showing slightly tighter bounds for IO. The LIV phases do not display independent sensitivity in our analysis, although they modulate the bounds on their associated coefficients around $\pi/2$ and $3\pi/2$.

Our analysis reveals that including the best-fit LIV parameters can modify the inferred oscillation parameter space. In particular, the NO best-fit points shift to higher values of the solar angle $\theta_{12}$ in the $c_{ee}$$-$$c_{e\tau}$ and $a_{ee}$$-$$a_{e\tau}$ scenarios. This shift reduces the overlap between the NO and IO allowed regions and leads to a clearer separation of their $3\sigma$ contours within the specific LIV configurations considered. This behavior, together with the lower $\chi^2_{\rm min}$ values obtained for IO in the $\sin^2\theta_{12}$--$\Delta m^2_{21}$ plane, suggests a preference for IO within our $\chi^2$ framework when the best-fit LIV parameters are included.

This result illustrates how subleading Lorentz-violating contributions can impact the relative ordering preference inferred from precision reactor data. The resulting reshuffling of the allowed parameter space may influence degeneracy patterns in global neutrino oscillation analyses. Together with the bounds obtained in this work, these findings highlight JUNO's potential to probe small deviations from standard neutrino propagation. With larger future datasets, JUNO is expected to significantly enhance sensitivity to isotropic Lorentz-violating effects as well as to the neutrino mass ordering.
\vspace{-1em}
\section*{Acknowledgments\label{sec:ack}}
This work was funded by the ANID FONDECYT/Regular 1241855. T.A. thanks the support from ANID-Chile through the National Doctoral Fellowship No. 21250478. S.P. acknowledges the funding from ANID-Chile under FONDECYT Postdoctorado No. 3250376. The authors acknowledge the discussions with Omar Miranda and Luis Delgadillo.

\clearpage
\appendix
\section{Results for Inverted Ordering}\label{secA1}

This appendix presents the corresponding inverted ordering results for the LIV parameter scans discussed in Section~\ref{sec:results}. In general, the IO allowed regions show a behavior similar to the NO case, with small changes in the location of the best-fit points and in the minimum $\chi^2$ values.

\begin{figure}[H]
    \centering
    \includegraphics[width=0.46\linewidth]{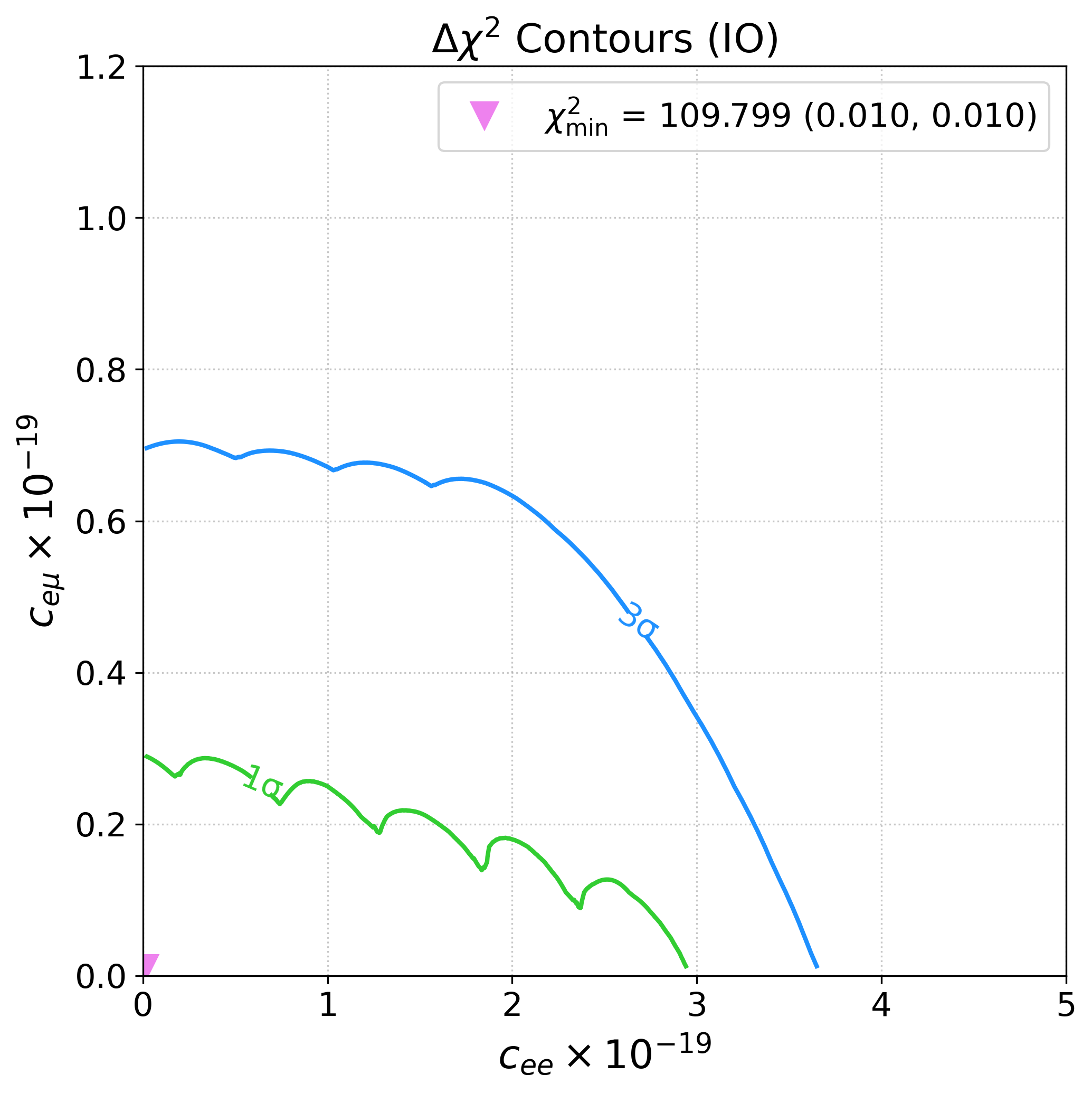}
    \includegraphics[width=0.46\linewidth]{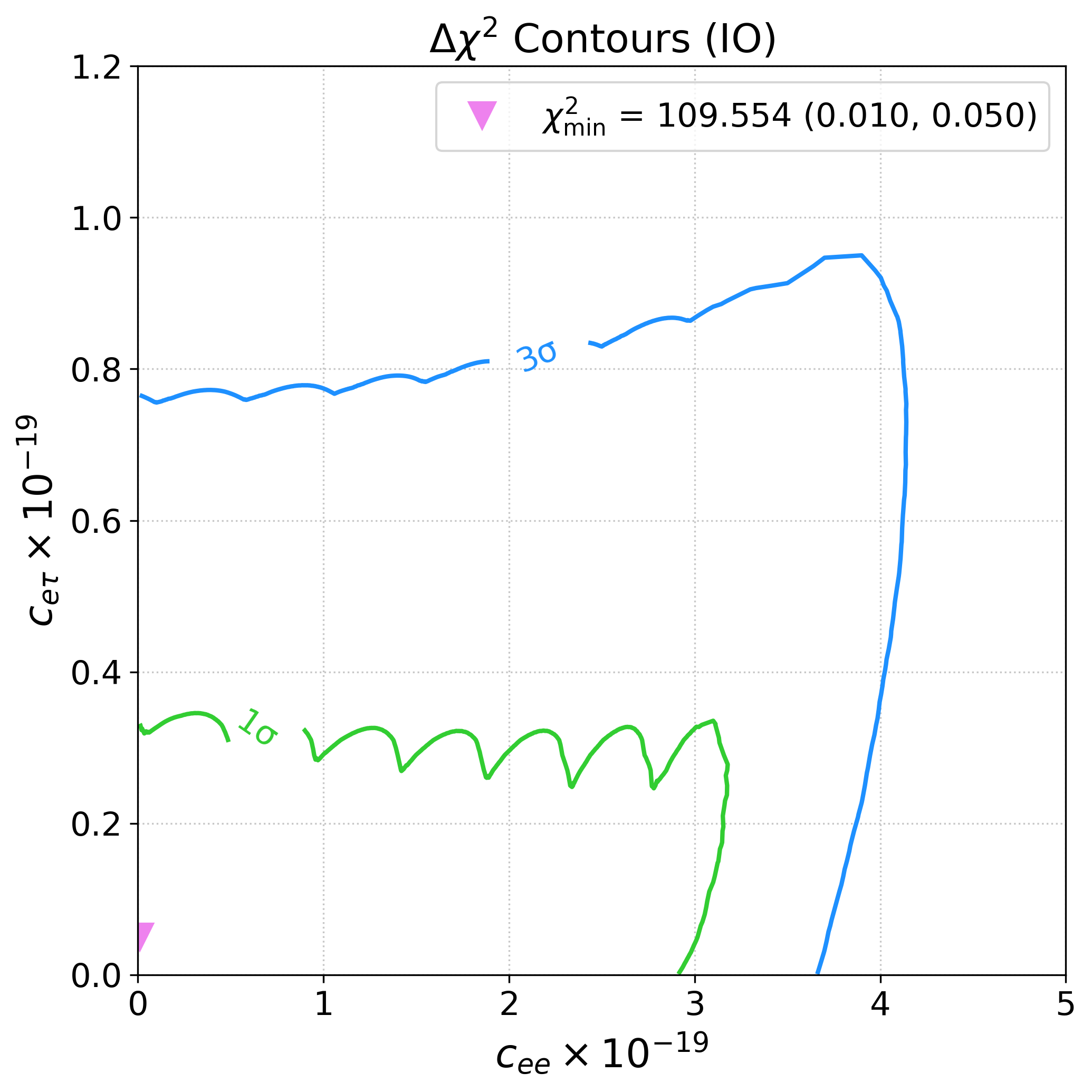}
    \caption{Two dimensional $\Delta\chi^{2}$ contours for the CP-conserving parameter pairs $(c_{ee}, c_{e\mu})$ (left) and $(c_{ee}, c_{e\tau})$ (right) assuming inverted ordering. The violet triangles mark the global minima of each scan, which correspond to the best-fit LIV values preferred by the data. The green and blue contours correspond to the $1\sigma$ and $3\sigma$ allowed regions, respectively.}
    \label{fig:chi-sqr-cee-emet-io}
\end{figure}

\begin{figure}[H]
    \centering
    \includegraphics[width=0.46\linewidth]{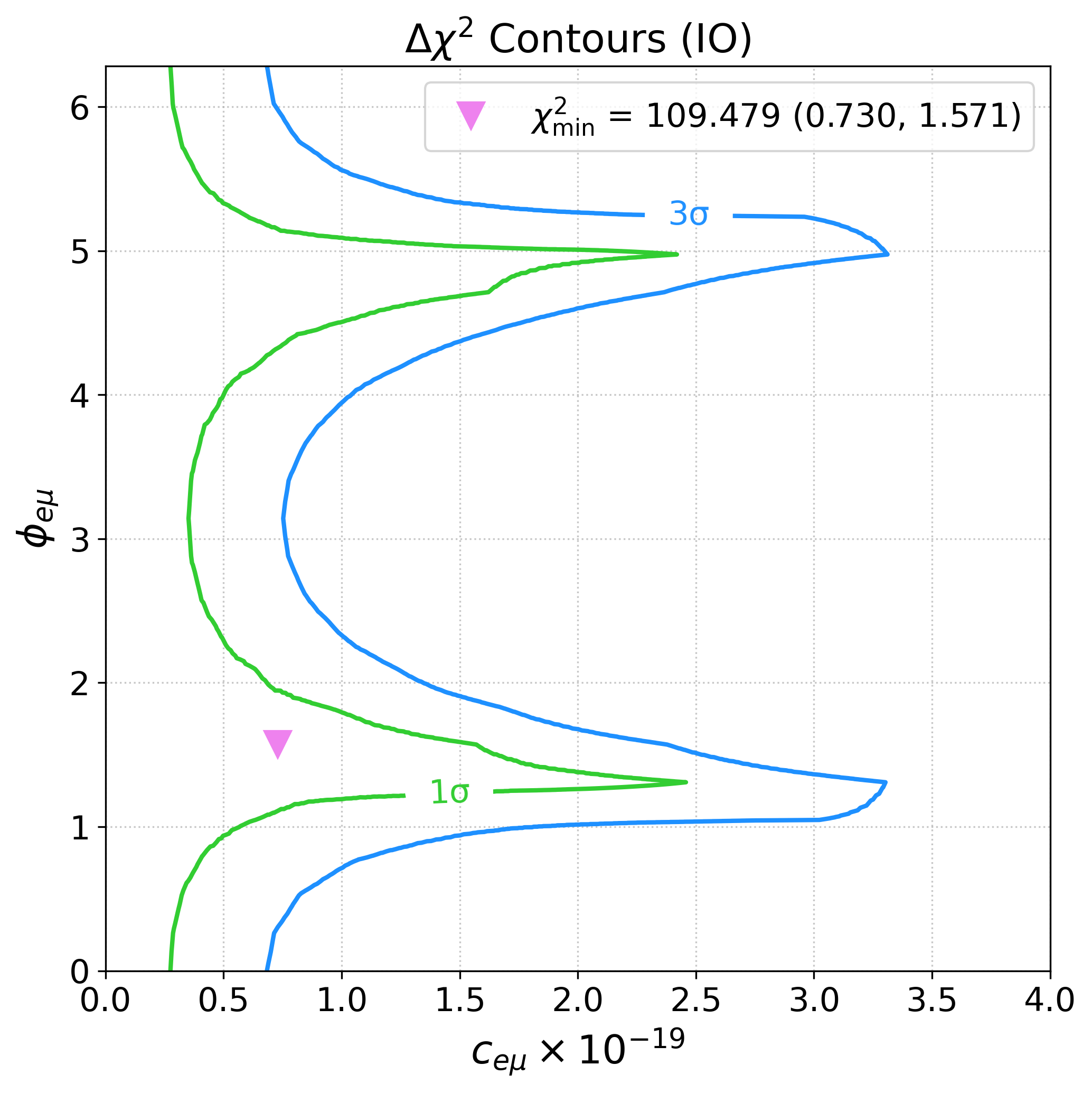}
    \includegraphics[width=0.46\linewidth]{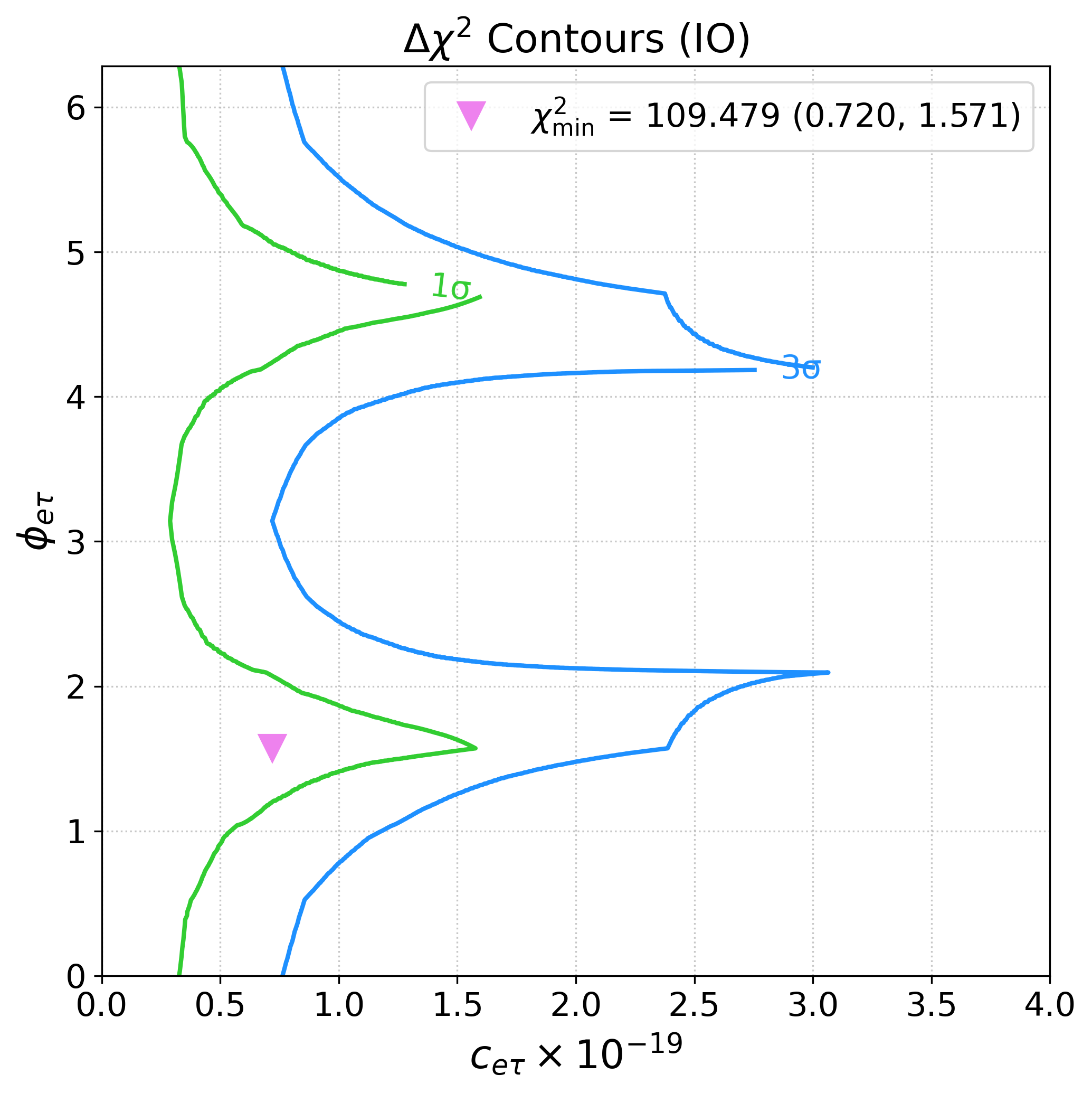}
    \caption{Two dimensional $\Delta\chi^{2}$ contours for the CP-conserving parameter pairs $(c_{e\mu}, \phi_{e\mu})$ (left) and $(c_{e\tau}, \phi_{e\tau})$ (right) assuming inverted ordering. The violet triangles mark the global minima of each scan, which correspond to the best-fit LIV values preferred by the data. The green and blue contours correspond to the $1\sigma$ and $3\sigma$ allowed regions, respectively.}
    \label{fig:chi-sqr-c-ph-io}
\end{figure}

Figure~\ref{fig:chi-sqr-cee-emet-io}, which can be compared with Figure~\ref{fig:chi-sqr-cee-emet-no}, shows the CPT-even scans in the $(c_{ee},c_{e\mu})$ and $(c_{ee},c_{e\tau})$ planes. The best-fit points are obtained at $(c_{ee},c_{e\mu})\simeq (0.01,0.01)\times10^{-19}$, with $\chi^2_{\min}=109.799$, and $(c_{ee},c_{e\tau})\simeq (0.01,0.05)\times10^{-19}$, with $\chi^2_{\min}=109.554$. These values are slightly lower than the standard oscillation value for IO, $\chi^2_{\min}=109.751$. The contours and best-fit points in Figure~\ref{fig:chi-sqr-c-ph-io}, obtained in the $(c_{e\mu},\phi_{e\mu})$ and $(c_{e\tau},\phi_{e\tau})$ planes, are also very similar to the corresponding NO results shown in Figure~\ref{fig:chi-sqr-c-ph-no}.

\begin{figure}[H]
    \centering
    \includegraphics[width=0.46\linewidth]{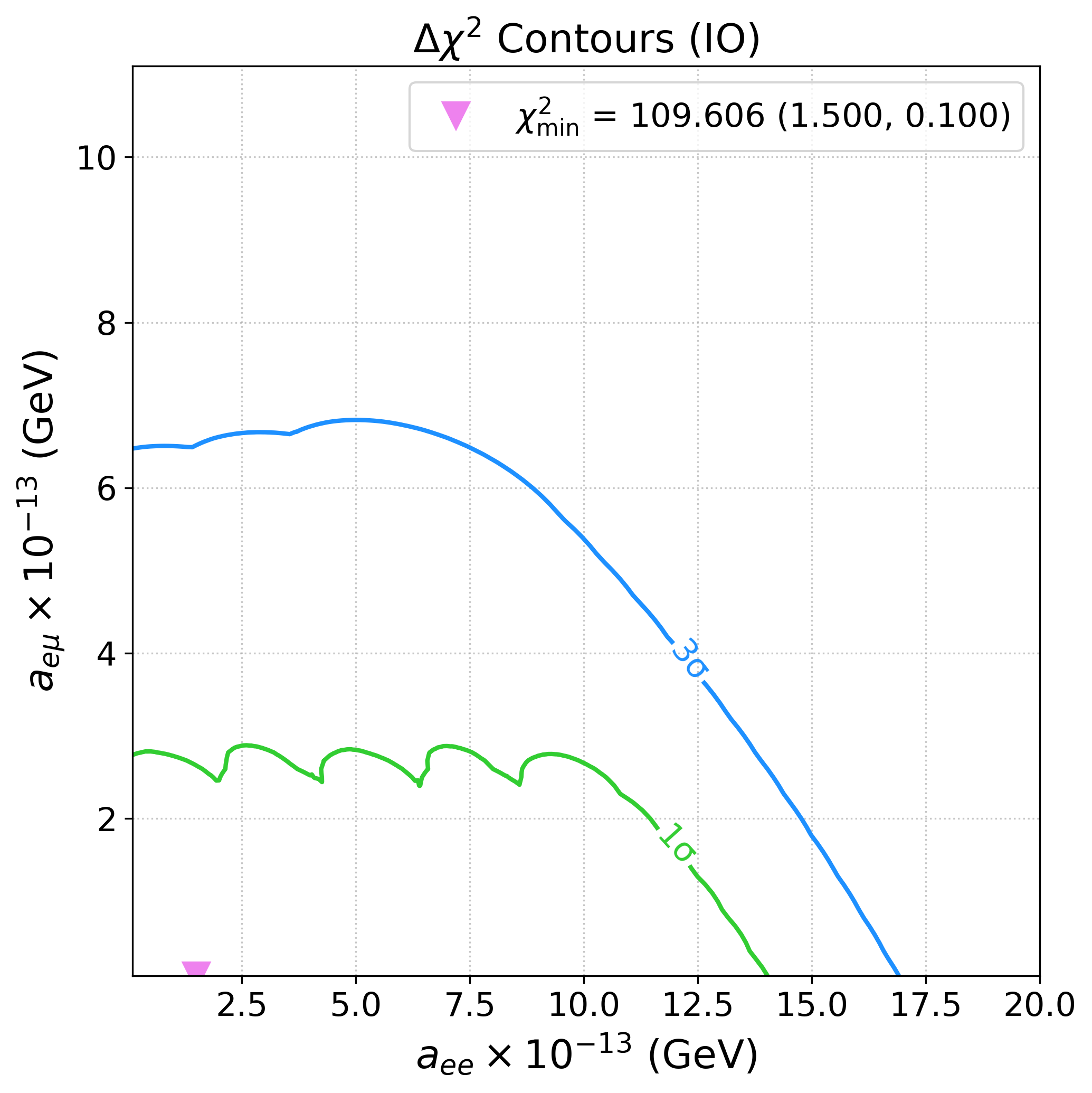}
    \includegraphics[width=0.46\linewidth]{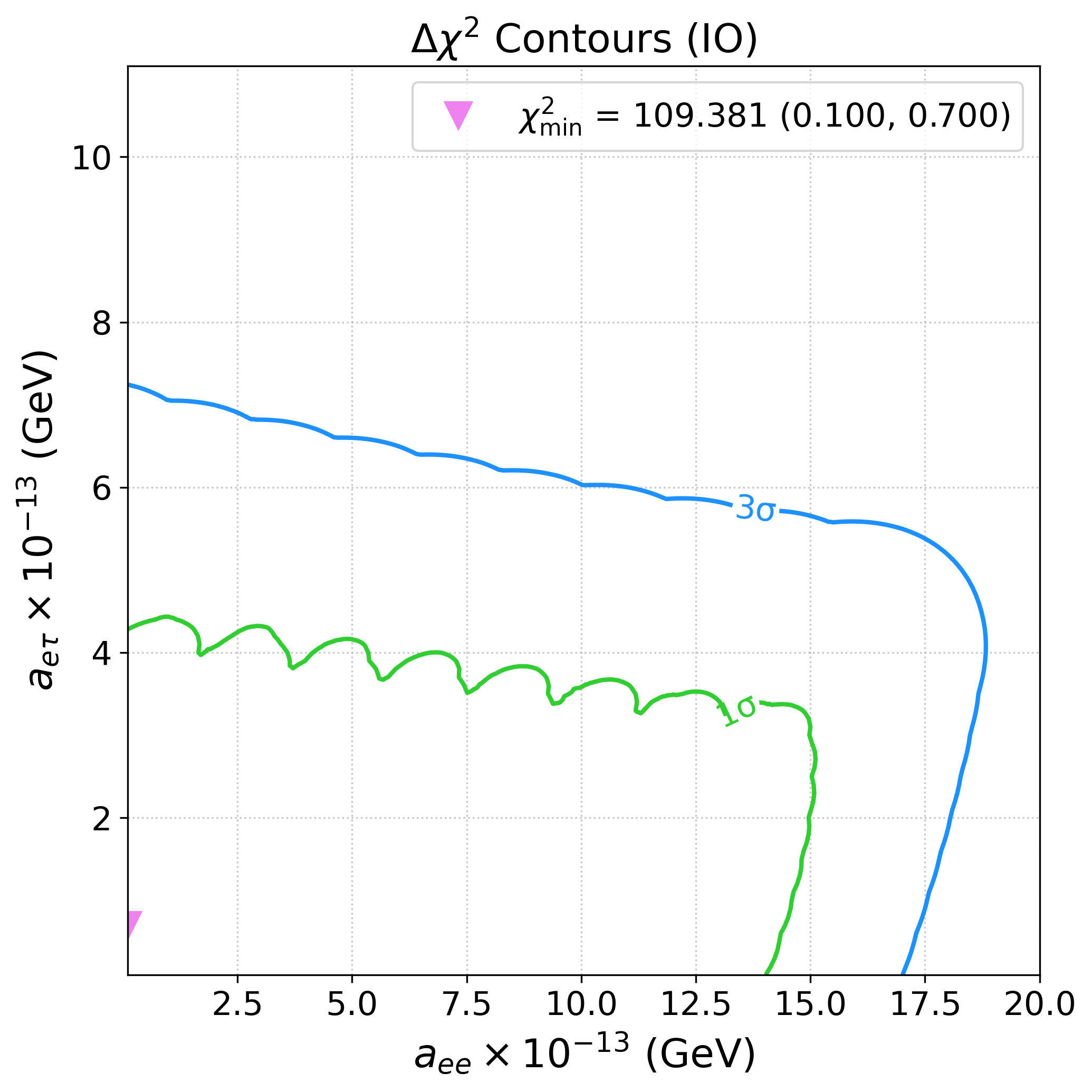}
    \caption{Two dimensional $\Delta\chi^{2}$ contours for the CP-violating parameter pairs $(a_{ee}, a_{e\mu})$ (left) and $(a_{ee}, a_{e\tau})$ (right) assuming inverted ordering. The violet triangles mark the global minima of each scan, which correspond to the best-fit LIV values preferred by the data. The green and blue contours correspond to the $1\sigma$ and $3\sigma$ allowed regions, respectively.}
    \label{fig:chi-sqr-aee-amet-io}
\end{figure}

For CP-violating cases, the $3\sigma$ regions shown in Figure \ref{fig:chi-sqr-aee-amet-io} for IO are also very similar to the NO case (Figure \ref{fig:chi-sqr-aee-amet-no}). The best-fit points are obtained at $(a_{ee},a_{e\mu})\simeq (1.5,0.1)\times10^{-13}$~GeV, with $\chi^2_{\min}=109.606$, and $(a_{ee},a_{e\tau})\simeq (0.1,0.7)\times10^{-13}$~GeV, with $\chi^2_{\min}=109.381$. These values are lower than the standard-oscillation value for IO, $\chi^2_{\min}=109.751$.

\begin{figure}[H]
    \centering
    \includegraphics[width=0.46\linewidth]{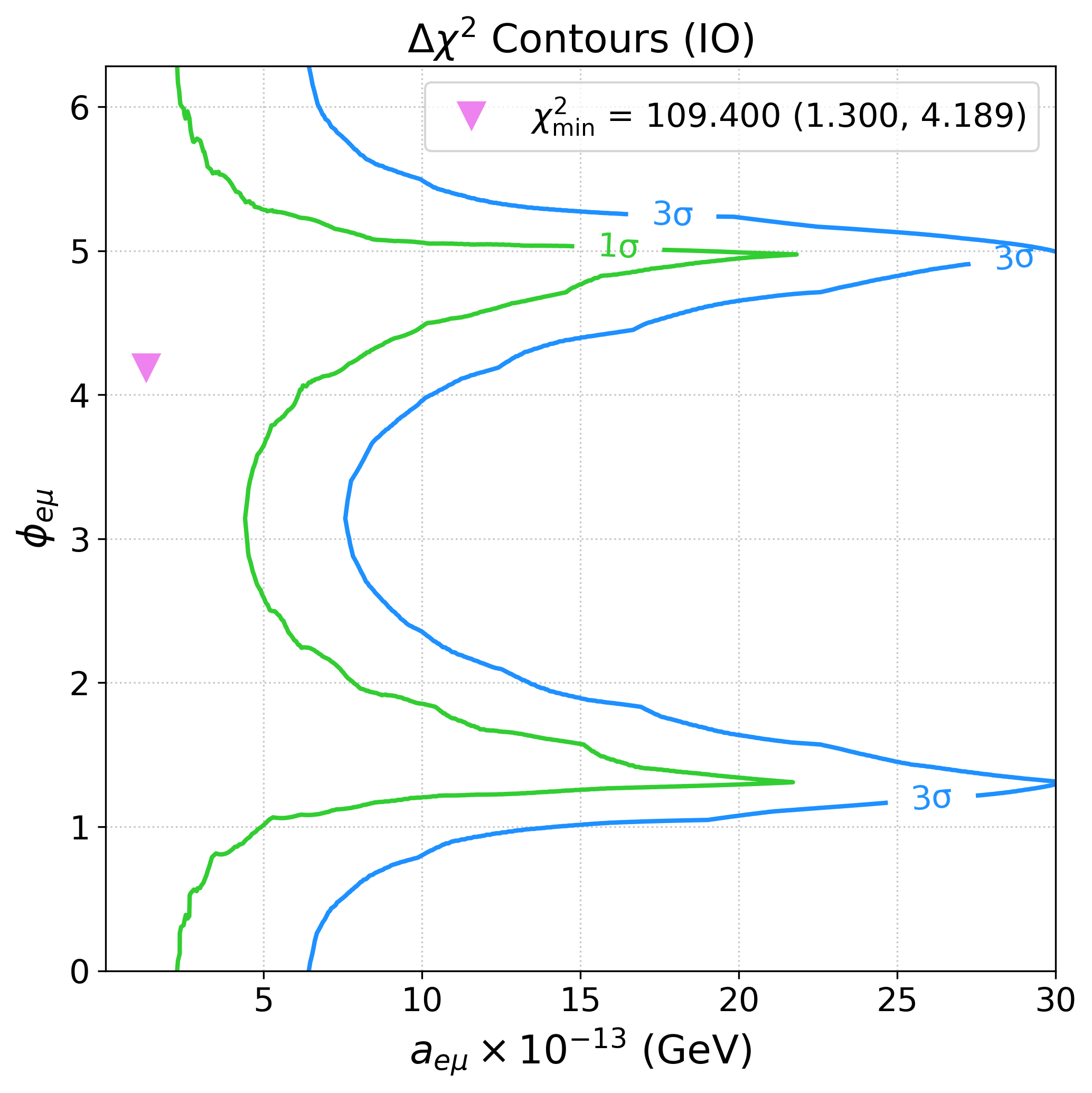}
    \includegraphics[width=0.46\linewidth]{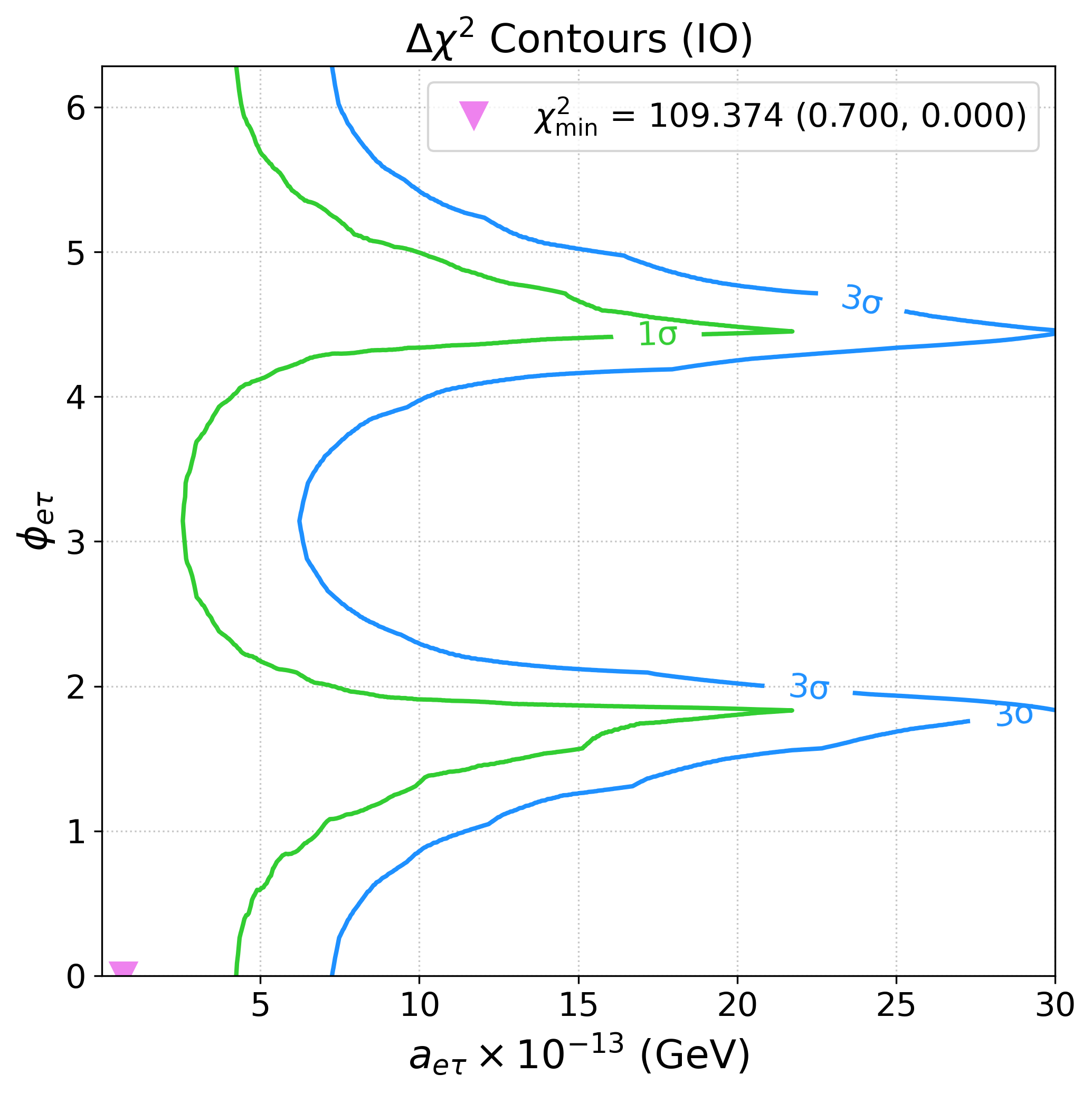}
    \caption{Two dimensional $\Delta\chi^{2}$ contours for the CP-violating parameter pairs $(a_{e\mu}, \phi_{e\mu})$ (left) and $(a_{e\tau}, \phi_{e\tau})$ (right) assuming inverted ordering. The violet triangles mark the global minima of each scan, which correspond to the best-fit LIV values preferred by the data. The green and blue contours correspond to the $1\sigma$ and $3\sigma$ allowed regions, respectively.}
    \label{fig:chi-sqr-a-ph-io}
\end{figure}

The contours in Figure \ref{fig:chi-sqr-a-ph-io}, corresponding to the $(a_{e\mu}, \phi_{e\mu})$ and $(a_{e\tau}, \phi_{e\tau})$ planes for IO, are mostly similar to the NO contours shown in Figure \ref{fig:chi-sqr-a-ph-no}. The main difference appears in the $(a_{e\mu},\phi_{e\mu})$ plane, where the IO best-fit value of $\phi_{e\mu}$ shifts to a larger value compared to the NO case. In the $(a_{e\tau},\phi_{e\tau})$ plane, the best-fit region remains similar for NO and IO.

\clearpage

\section{Supplementary SM--LIV comparisons at JUNO}
\label{sec:SM-LIV-comparison}

In this appendix, we illustrate the effect of the LIV best-fit parameters at the event-spectrum level as a function of prompt energy and compare with the SM best-fit. The LIV coefficients are fixed to the best-fit values obtained from the $\chi^2$ scans and listed in Tables~\ref{tab:cpc-no} and~\ref{tab:cpc-io}. 

\begin{figure}[h!]
    \centering
    \includegraphics[width=0.46\linewidth]{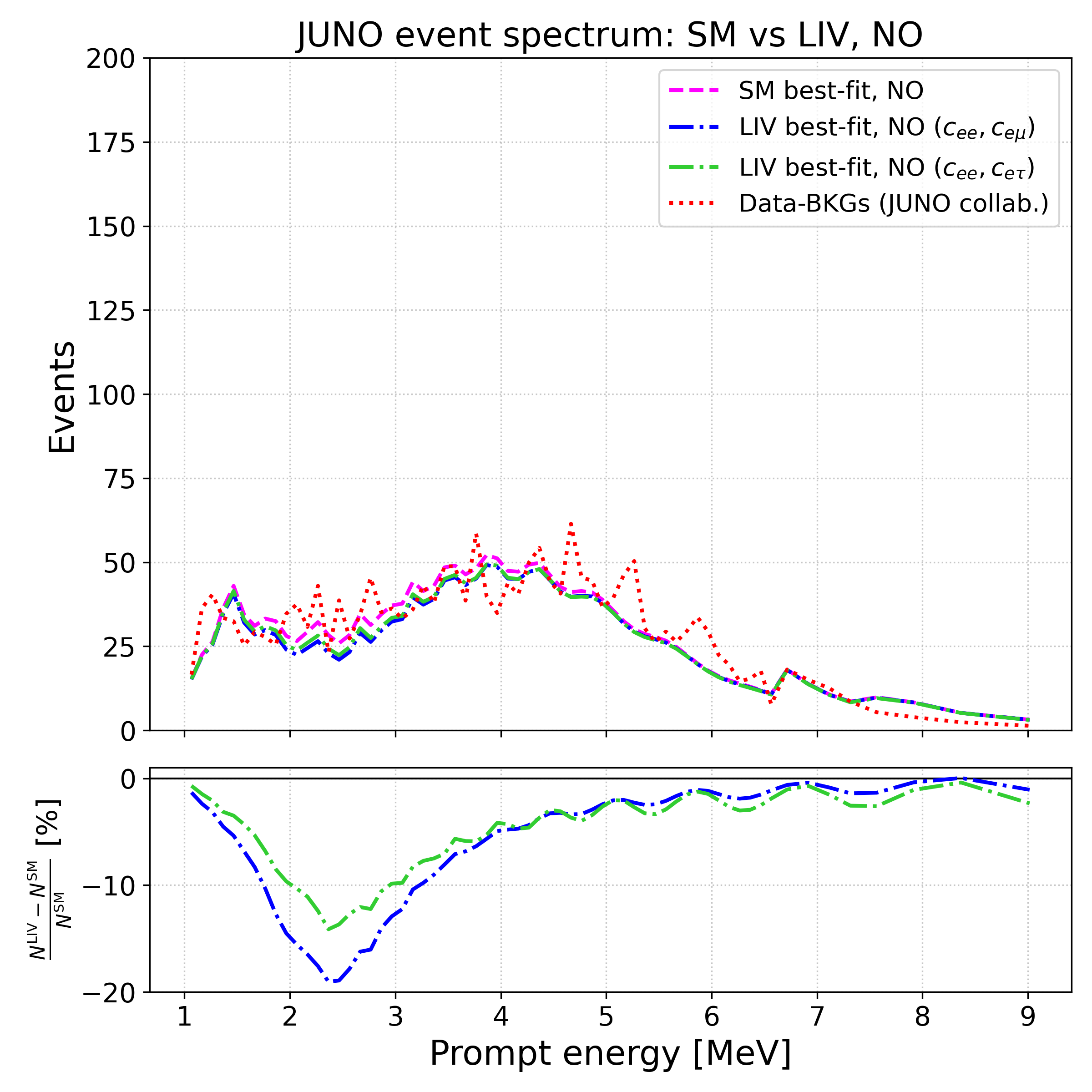}
    \includegraphics[width=0.46\linewidth]{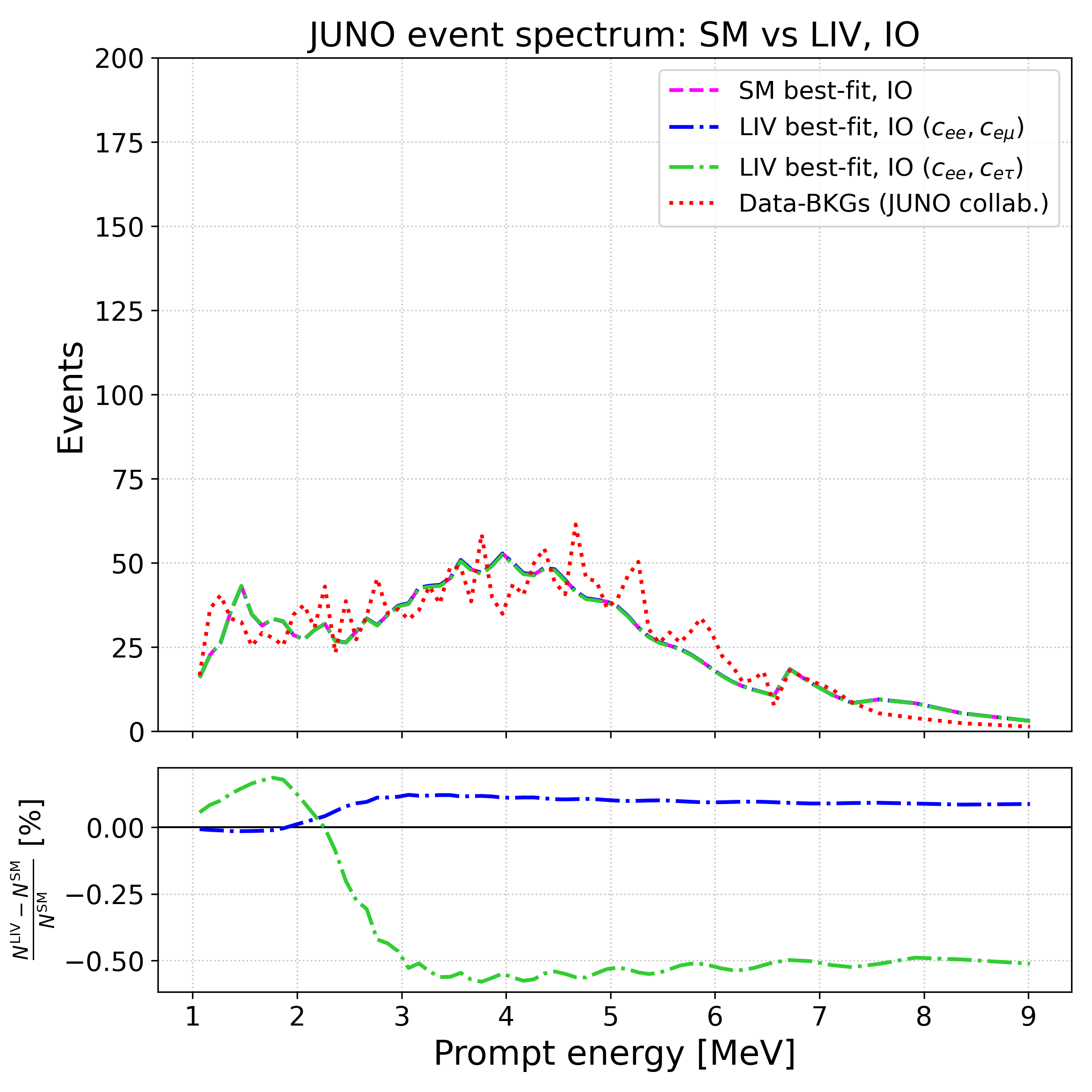}
    \caption{Comparison of the reconstructed JUNO event spectrum in the standard-oscillation scenario and in representative LIV best-fit scenarios. The left and right panels correspond to normal and inverted ordering, respectively. In each panel, the upper plot shows the SM best-fit spectrum, the LIV best-fit spectra for the $(c_{ee},c_{e\mu})$ and $(c_{ee},c_{e\tau})$ configurations, and the JUNO data after background subtraction (JUNO Collaboration) \cite{abusleme2025first}. The lower plot shows the relative deviation of each LIV prediction with respect to the SM best-fit prediction, $(N_i^{\rm LIV}-N_i^{\rm SM})/N_i^{\rm SM}$. These residuals illustrate the small energy-dependent distortions induced by the LIV coefficients.}
    \label{fig:event-spectrum-sm-liv}
\end{figure}

Figure~\ref{fig:event-spectrum-sm-liv} compares the JUNO event-rate spectrum predicted in the SM scenario  with the corresponding isotropic LIV spectra for the $c_{ee}$$-$$c_{e\mu}$, and $c_{ee}$$-$$c_{e\tau}$ best configurations. The left and right panels correspond to normal ordering and inverted ordering, respectively. In each panel, the red dotted curve represents the JUNO data after background subtraction.

The event-spectra with LIV best-fit values remain close to the SM predictions, especially for inverted ordering. In the isotropic LIV scenarios considered here, the effect does not appear as a localized excess in the reconstructed prompt-energy spectrum, instead, it appears as a small energy-dependent distortion. The ordering preference discussed in the main text is quantified through the evaluation of $\chi^2_{\rm min}$, and not by visual inspection of the event spectra alone.

\clearpage
\printbibliography
\end{document}